\documentclass[english]{article}
\usepackage[T1]{fontenc}
\usepackage[latin9]{inputenc}
\usepackage{geometry}
\usepackage{color}
\usepackage{float}
\usepackage{textcomp}
\usepackage{mathtools}
\usepackage{amsmath}
\usepackage{amssymb}
\usepackage{graphicx}

\makeatletter

\providecommand{\tabularnewline}{\\}

\numberwithin{equation}{section}
\numberwithin{figure}{section}
\newcommand{\lyxaddress}[1]{
	\par {\raggedright #1
	\vspace{1.4em}
	\noindent\par}
}

\@ifundefined{showcaptionsetup}{}{%
 \PassOptionsToPackage{caption=false}{subfig}}
\usepackage{subfig}
\makeatother

\usepackage{babel}
\begin{document}
\title{Mathematical modeling for sustainable aphid control in agriculture
via intercropping}
\author{Alfonso Allen-Perkins$^{1,2}$, Ernesto Estrada$^{3,4,5}${*}}
\maketitle

\lyxaddress{$^{1}$Instituto de F\'isica, Universidade Federal da Bahia, 40210-210
Salvador, Brazil; $^{2}$Complex System Group, Universidad Polit\'ecnica
de Madrid, 28040-Madrid, Spain; $^{3}$Institute of Applied Mathematics
(IUMA), Universidad de Zaragoza, Pedro Cerbuna 12, E-50009 Zaragoza,
Spain; $^{4}$ARAID Foundation, Government of Arag\'on, 50018 Zaragoza,
Spain; $^{5}$Instituto de Ci\^encias Matem\'aticas e de Computa\c{c}\~ao, Universidade
de S\~ao Paulo, Caixa Postal 668, 13560-970 S\~ao Carlos, S\~ao Paulo, Brazil.}

{*}Corresponding author: Ernesto Estrada, email: estrada66@unizar.es
\begin{abstract}
\textcolor{black}{Agricultural losses to pest represent an important
challenge in a global warming scenario. Intercropping is an alternative
farming practice that promotes pest control without the use of chemical
pesticides. Here we develop a mathematical model }to study epidemic
spreading and control in intercropped agricultural fields as a sustainable
pest management tool for agriculture. The model combines the movement
of aphids transmitting a virus in an agricultural field, the spatial
distribution of plants in the intercropped field, and the presence
of ``trap crops'' in an epidemiological Susceptible-Infected-Removed
(SIR) model. Using this model we study several intercropping arrangements
without and with trap crops and find a new intercropping arrangement
that may improve significantly pest management in agricultural fields
respect to the commonly used intercrop systems.
\end{abstract}

\section{Introduction}

\textcolor{black}{The sustainable intensification of agriculture is
imperative for feeding a growing world population while minimizing
its negative environmental impact. }The world population will increase
to between 9.6 and 12.3 billion in 2100 \cite{World population},
and for feeding these additional 2-4 billion people, a duplication
(100-110\%) of crop production relative to its 2005 level is needed
\cite{food demand}. Today, 10\% of ice-free land on Earth is used
for crop cultivation \cite{Production duplication}, and returning
half of Earth\textquoteright s terrestrial ecoregions to nature will
mean global losses of 15\textendash 31\% of cropland and of 3\textendash 29\%
of food calories \cite{challenge}. Thus, increasing crop yield without
extending the size of cultivation areas nor by intensifying the use
of current technologies is a vital complex problem to be solved in
the coming years. Agricultural yield is substantially reduced by pests
\cite{pest_losses_1,pest_losses_2,pest_losses_3,pest_losses_4,pest_losses_5},
which cause losses of 10-16\% to crop production \cite{pest_losses_1,pest_losses_2,pest_losses_3,pest_losses_4,pest_losses_5},
which may represent real \textcolor{black}{threats} for entire world
regions \cite{Africa}. In addition to these scenarios, there is increasing
concern that climate change can increase plant damage from pests in
future decades \cite{climate_1,climate_2,climate_3,climate_4,climate_5,climate_6}.
Bebber et al. \cite{Bebber} have demonstrated that pests and pathogens
have shifted poleward by $2.7\pm0.8$ km/yr since 1960. This will
produce lower numerical response of biological control agents, which
can be translated into higher probabilities of insect pest outbreaks.
Deutsch et al. \cite{Deutsch} estimated that global yield losses
of rice, maize and wheat grains are projected to increase in the range
of 10 to 25\% per degree of global mean surface warming. Thus, in
a projected scenario of 2$^{\circ}$C-warmer climate the mean increase in yield
losses owing only to pest pressure extend to 59, 92, and 62 metric
megatons per year for wheat, rice and maize, respectively \cite{Deutsch}.
These losses cover most of the globe as can be seen in the Fig. 2
in ref. \cite{Deutsch}, but they are primarily centered in temperate
regions. 

From the agricultural point of view, a particularly important class
of insect pests are the aphids (aphididae) \cite{Aphids book}. Aphids
are by far the most important transmissors of plant viruses, being
reported to transmit about 50\% of insect-borne plant viruses (approximately
275 virus species). There are about 4,700 aphids described from which
about 190 transmit plant viruses (see Chapter 15 of \cite{Aphids book}).
From the economic point of view this virus transmission by aphid represents
global losses estimated on tens of millions to billions US\$ of yield
loss per annum \cite{aphids economics 1,aphids economics 2,aphids economics 3}.
In the UK alone the damage on cereals made by aphids has been estimated
to be around 60-120 million pounds annually \cite{aphids economics 4}.
Thus, mathematical modeling is seen as an important tool to predict
and mitigate the effects of viruses on agriculture \cite{Kenya,Cocoa}. 

Today, there are several alternative approaches for the sustainable
intensification of agriculture based on agroecological and adaptive
management techniques \cite{Agroecological}. A recent work reports
evidences that organic farming, for instance, promotes pest control
\cite{Organic}. An example is\textit{ intercropping,} consisting
in growing two or more crops in the same field, which has proved to
be important for pest control in several crops \cite{Khan_et_al,intercropping_1,Martin-Guay}
(see Supplementary Table 1). Intercropping is known since the 16th-18th
centuries when Iroquoian farmers inter-planted the \textit{Three Sisters}:
corn, bean, and squash \cite{Iroquoain}. Intercropping is known to
reduce the levels of infestation by stemborers and increases insect
pest parasitism \cite{parasitism}. These practices have been extended
across the globe as can be seen in Fig. 1 of ref. \cite{Martin-Guay}.
Meta-analysis of 552 experiments in 45 papers published between 1998
and 2008 \textcolor{black}{showed} that intercropping produces significant
improvement for herbivore suppression, enemy enhancement, and crop
damage suppression \cite{meta_analysis} respect to monocrop. Brooker
et al. \cite{Brooker} have concluded that intercropping ``\textit{could
be one route to delivering 'sustainable intensification}'{}'' of
agriculture. In the particular case of aphids, there are many reports
on the successful use of intercropping strategies for controlling
aphid-transmitted viral diseases \cite{aphids intercropping 1,aphid intercropping 2,aphid intercropping 3}.
In a recent review, a series of companion plants that can be potentially
used in intercropping strategies for controlling aphids have been
reported, together with several strategies for controlling aphid-produced
diseases \cite{aphid intercropping 4}. 

Here we develop and implement a mathematical model that allow us to
study intercropping as a sustainable pest management tool for agriculture.
Our main goal is to investigate which are the best spatial arrangements
for controlling aphid-transmitted viruses in agricultural scenarios
by avoiding the propagation of aphids through the crop field. For
this purpose we combine the movement of aphids in the agricultural
landscape \cite{insect movement 1,insect movement 1_1,insect movement 2,insect movement 3}
with the spatial distribution of plants in the intercropped field,
in an epidemiological Susceptible-Infected-Removed (SIR) \cite{Infectious}
model. The model allows us to implement ``trap crops''\textendash plants
which attract or detract insects to protect target crops \cite{trap_crop_1,trap_crop_2,trap_crop_3,trap_crop_4,trap_crop_5}.
Using this approach we find that a new intercropping arrangement proposed
here\textendash particularly when combined with trap crops\textendash can
improve significantly pest management in agricultural fields respect
to the commonly used intercrop systems.

\section{Theoretical Methods}

For the development of the theoretical model to be used in this work
we make the following assumptions:
\begin{enumerate}
\item The infection is transmitted to plants by an aphid\textendash a vector.
That is, a susceptible plant receives the infection, e.g., a virus,
from an infectious plant through a vector.
\item Recovered (removed) plants represent those not only dead but also
those which are useless for commercial purposes, i.e., those substantially
damaged as to be used for consumption.
\item The number of plants in the field is fixed.
\item When a susceptible vector is infected by a plant, there is a fixed
time $\tau$ during which the infectious agent develops in the vector.
At the end of this time, the vector can transmit the virus to a susceptible
plant.
\item The number of infectious vectors is very large and at a given time
$t$ its amount is proportional to $I\left(t-\tau\right)$.
\end{enumerate}
These assumptions are an adaptation of the ones made by Cooke \cite{Cooke}
for implementing a time-delay Susceptible-Infected-Recovered (SIR)
model to study a vector-borne infection transmission to a given population.
The corresponding equations read as follows:

\begin{equation}
\begin{split}\dot{S}_{i}\left(t\right) & =-\beta S_{i}\left(t\right)\sum_{j}I_{j}\left(t-\tau\right),\\
\dot{I}_{i}\left(t\right) & =\beta S_{i}\left(t\right)\sum_{j}I_{j}\left(t-\tau\right)-\mu I_{i}\left(t\right),\\
\dot{R}_{i} & \left(t\right)=\mu I_{i}\left(t\right),
\end{split}
\end{equation}
where $S_{i}$ is the probability of plant $i$ of being susceptible
to the infection, $I_{i}$ is the probability of plant $i$ of being
infective after having been infected by the disease, and $R_{i}$
is the probability of plant $i$ of being removed, $\beta$ and $\mu$,
are the birth and death rates of the disease, respectively,\textcolor{black}{{}
and $j$ spans only to the plants that are able to spread the disease
by contact to plant $i$. Note that $S_{i}\left(t\right)+I_{i}\left(t\right)+R_{i}\left(t\right)=1$,
and, consequently, $\dot{S}_{i}\left(t\right)+\dot{I}_{i}\left(t\right)+\dot{R}_{i}\left(t\right)=0$.
}This model has been subsequently studied in the literature by several
authors as a vector-borne disease transmission model (see for instance
\cite{Cooke_2,Cooke_3,Cooke_4,Cooke _5}). For other approaches to
modeling vector-borne virus transmission on plants see for instance
\cite{Virus modeling 2}.

Here we generalize Cooke's model \cite{Cooke} in order to account
for the probability that a vector hops not only to a neighboring plant
but also to a more distant one in the field:

\begin{equation}
\dot{S}_{i}\left(t\right)=-\beta S_{i}\left(t\right)\sum_{j}f_{ij}I_{j}\left(t-\tau\right),\label{eq1-1}
\end{equation}

\begin{equation}
\dot{I}_{i}\left(t\right)=\beta S_{i}\left(t\right)\sum_{j}f_{ij}I_{j}\left(t-\tau\right)-\mu I_{i}\left(t\right),\label{eq2-1}
\end{equation}

\begin{equation}
\dot{R}_{i}\left(t\right)=\mu I_{i}\left(t\right),
\end{equation}
where $f_{ij}$ is a function of the ``separation'' between the
plants $i$ and $j$, \textcolor{black}{and $j$ spans to all the
plants in the field. There are two possibilities of accounting for
this separation between plants. The first is to consider the Euclidean
distance between the corresponding two plants, i.e., $\rho_{ij}=\sqrt{\left(x_{i}-x_{j}\right)^{2}+\left(y_{i}-y_{j}\right)^{2}}$,
where $x_{i}$ and $y_{i}$ are the Cartesian coordinates of the plant
$i$ in the plane. Notice that this distance is not capturing all
the subtleties of the real separation between the plants as two plants
can be of different height, and a third coordinate should be introduced.
In this case we can consider that the probability $f_{ij}$ of moving
from plant $i$ to plant $j$ is proportional to certain function
of this distance, e.g., decaying as a power-law $f_{ij}\propto\rho_{ij}^{-s}$
or decaying exponentially $f_{ij}\propto\exp\left(-\lambda\rho_{ij}\right)$,
where $s,\lambda\in\mathbb{R}^{+}$. }

\textcolor{black}{The second approach is to consider the plant-to-plant
separation in terms of the number of hops that an aphid needs to take
to go from plant $i$ to plant $j$ using other intermediate plants.
That is, let us consider that the aphid in question has an exploration
radius equal to $r$. This means that if the aphid is on plant $i$
it can hop directly to a plant which is at a distance smaller than
$r$ from $i$. In order to hop to a plant $k$ separated by two radii
from $i$ it has to use two steps. That is, if we connect two plants
by an edge if their geographic separation is $\rho_{ij}\leq r$, then
the plant-to-plant (topological) separation $d_{ij}$ is given by
the number of edges in the shortest path connecting the two nodes
in the resulting graph $G=\left(V,E\right)$ with vertices $V$ and
edges $E$. In this case we again can consider that the probability
$f_{ij}$ of moving from plant $i$ to plant $j$ is proportional
to certain function of this distance, e.g., decaying as a power-law
$f_{ij}\propto d_{ij}^{-s}$ or decaying exponential}ly $f_{ij}\propto\exp\left(-\lambda d_{ij}\right)$,
where $s,\lambda\in\mathbb{R}^{+}$. Let us consider some of the potential
differences between these two ways of accounting for the interplant
separation.

\subsection{The rationale of the model: Through-space vs. plant-to-plant aphid
mobility}

\textcolor{black}{From the complex movements that an aphid can display
in a crop field (see Chapter 10 in \cite{Aphids book} and \cite{behavioural,flight}),
here we focus only on their exploratory movement inside a crop field.
This includes mainly displacements to neighboring plant (primary movement)
or a distant plant inside the same field. We exclude from here those
unintentional movements of aphids such as the displacement by air
currents that can transport them at very long geographic distances.
Thus, with this restriction in mind we analyze the main differences
in considering a model that includes geographic or topological distance
for epidemic transmission. In doing so, we have identified three main
factors in favor of the use of the topological interplant separation
which are based on the main behavioral characteristics of aphids exploratory
movement inside crop fields \cite{Aphids-plants 1,Aphids-plants 2,Aphids-plants 3,Aphids-plants 4}.
These three principles are the following: (i) }\textbf{\textit{\textcolor{black}{first
come first served}}}\textcolor{black}{, which essentially tells that
an aphid flying in one direction will land in the first available
plant independently of the distance at which it is from its starting
position; (ii) }\textbf{\textit{\textcolor{black}{a bird in the hand
is worth more than two in the bush}}}\textcolor{black}{, indicating
that the probability that an aphid moves from a plant $i$ to another
$j$ decays with the number of other plants in the path between $i$
and $j$; (iii) }\textbf{\textit{\textcolor{black}{go back before
it is too late}}}\textcolor{black}{, which indicates that an aphid
flying in a direction without plants would prefer to return to its
starting point. These principles are detailed in the Supplementary
Note 1.}

\subsection{SIR model with topological distances}

\begin{figure}[H]
\begin{centering}
\includegraphics[width=0.45\textwidth]{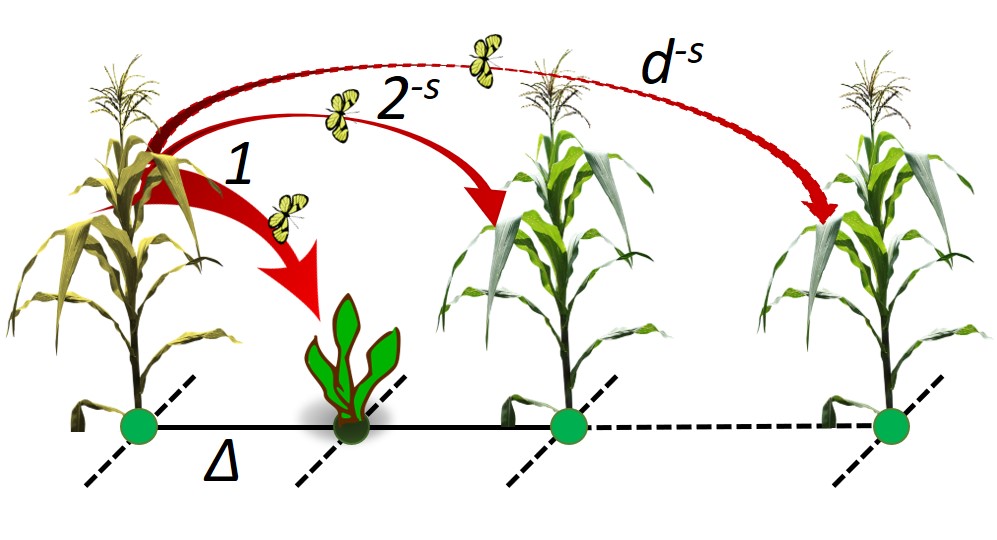}
\par\end{centering}
\caption{Inter-plants movements of an aphid in an agricultural plot with intercropping
(see Supplementary Note 1). The hop of an aphid from an infected plant
to a susceptible one separated by $d$ steps is given by $d^{-s}$
(see Supplementary Note 1). }

\label{hopping}
\end{figure}

As a consequence of the previous hypothesis we conclude that the use
of the topological interplant separation is appropriate for our modeling
purposes. Therefore, the SIR model on the field is expressed as \cite{LRI_plants}:

\begin{equation}
\dot{S}_{i}\left(t\right)=-\beta S_{i}\sum_{j}\tilde{A}_{ij}I_{j}\left(t-\tau\right),\label{eq1}
\end{equation}

\begin{equation}
\dot{I}_{i}\left(t\right)=\beta S_{i}\left(t\right)\sum_{j}\tilde{A}_{ij}I_{j}\left(t-\tau\right)-\mu I_{i}\left(t\right),\label{eq2}
\end{equation}

\begin{equation}
\dot{R}_{i}\left(t\right)=\mu I_{i}\left(t\right),
\end{equation}
where $\tilde{A}=\sum_{d=1}^{d_{max}}d^{-s}A_{d}$ , $d\leq d_{max}$,
$d_{max}$ is diameter of the graph, i.e., the largest separation
between two plants (in terms of steps), and the matrix $A_{d}$ captures
the (long-range) mobility of the pest between plants (see Fig. \ref{hopping}).

\textcolor{black}{The $d$-path adjacency matrices $A_{d}$ used in
the current formulation are generalizations of the concept of adjacency
matrix. In the Supplementary Note 2 we give a formal definition of
them and an example (see also \cite{path Laplacian 1,path Laplacian 2,path Laplacian 3}).
We will always consider a connected network here. Let $d_{ij}=d$
be the shortest-path distance between the nodes $i$ and $j$ in a
network $G$. Then, the $d$-path adjacency matrix is defined by}

\textcolor{black}{
\begin{equation}
A_{d}\left(i,j\right)=\left\{ \begin{array}{cc}
1 & \textnormal{if }d_{ij}=d\\
0 & \textnormal{otherwise}.
\end{array}\right.
\end{equation}
}

\textcolor{black}{We consider any $1\leq d\leq d_{max}$, where $d=1$
provides the ``classical'' adjacency matrix and where $d_{max}$
is the diameter of the network. Then, we combine all the $d$-path
adjacency matrices by using a transformation, such that}

\textcolor{black}{
\begin{equation}
\tilde{A}=A_{1}+2^{-s}A_{2}+\cdots+d_{max}^{-s}A_{d_{max}},\label{eq:Mellin}
\end{equation}
where $s$ is an empirical parameter controlling the insect mobility.
The transformation in Eq. (\ref{eq:Mellin}) is denoted as Mellin
$d$-path transformation. In this case, the entries of $\tilde{A}$
are defined as follow:}

\textcolor{black}{
\begin{equation}
\widetilde{A}\left(i,j\right)=\left\{ \begin{array}{cc}
d_{ij}^{-s} & \textnormal{if }i\neq j\\
0 & \textnormal{if }i=j.
\end{array}\right.
\end{equation}
}

Notice that the transformed adjacency matrix $\tilde{A}$ is symmetric
in the case of undirected networks. Then, when the aphid has very
poor mobility $s\rightarrow\infty$, all the entries of $\tilde{A}$,
except those equal to one, become zeroes, which indicates that the
aphid can only perform hops to nearest neighbors. On the other hand,
when the aphid has a very large mobility $s\rightarrow0$, every entry
of $\tilde{A}$ becomes one, which means that the aphid can hop from
one plant to another with equal probability independently of their
separation in the field.

\subsection{\textcolor{black}{Markovian formulation of the epidemiological model.}}

\textcolor{black}{Following the framework introduced in \cite{Epidemics_4},
we formulate a Markovian dynamics that, in principle, is valid for
any epidemic prevalence. For that reason, hereafter, we restrict ourselves
to this Markovian approach. Let $p_{i}(t)$ be the probability that
a node $i$ is infected at time $t$. Then, in the SIR model, the
Markovian equations reads as follows: 
\begin{eqnarray}
p_{i}(t+1) & = & p_{i}(t)(1-\mu)+(1-p_{i}(t)-\varrho_{i}(t))q_{i}(t-\tau)\;,\label{sir1}\\
\varrho_{i}(t+1) & = & \varrho_{i}(t)+\mu p_{i}(t)\;,\label{sir2}
\end{eqnarray}
where $\varrho_{i}(t)$ is the probability that node $i$ is removed
at time $t$. Note that the term $1-p_{i}(t)-\varrho_{i}(t)$ is just
$s_{i}(t)$, the probability that a node $i$ is susceptible at time
$t$. The expression for the infection probability $q_{i}(t-\tau)$
is \cite{LRI_plants} }

\textcolor{black}{
\begin{equation}
q_{i}(t-\tau)=1-\prod_{j=1}^{N}\left[1-\beta\tilde{A}_{ij}p_{j}(t-\tau)\right],\label{infect}
\end{equation}
which represents the probability that, when node $i$ is healthy at
time $t$, it becomes infected at time $t+1$. The expression $q_{i}$
is calculated as $1$ minus the probability that the node $i$ is
not infected by any infectious contact. This last probability is the
product over all the possible contacts of node $i$, considering that
a node $j$ transmits the disease to $i$ with probability $\beta\tilde{A}_{ij}p_{j}$,
after the delay time $\tau$. Note that if node $j$ is not connected
to $i$ (i.e., if $d_{ij}>1$ and $s\rightarrow\infty$), $\tilde{A}_{ij}=0$,
then the corresponding term in the product is equal to $1$, since
$j$ cannot infect $i$ regardless of its state, $p_{j}(t-\tau)$.}

\textcolor{black}{We should notice that this Markovian formulation
holds for any disease incidence, while Eqs. (\ref{eq1}) and (\ref{eq2})
are only valid when the disease prevalence is small. To explain this,
take Eq. (\ref{infect}) for $q_{i}(t-\tau)$ and consider that the
prevalence is small, $p_{i}\ll1$ $\forall i$, and for this reason
let us denote $p_{i}=x_{i}$. Then, the product in (\ref{infect})
transforms into: $1-\sum_{j=1}^{N}\beta\tilde{A}_{ij}x_{j}$. the
new expression for $q_{i}(t)$ in Eq. (\ref{sir1}), and passing from
discrete to continuous time, we recover a similar expression to that
in Eq. (\ref{eq2}) for the evolution of the infected state of node
$i$. For more details the reader is referred to \cite{details}.}

\textcolor{black}{The rate of propagation of the aphid-borne viral
infection across an agricultural field is defined here as}

\textcolor{black}{
\begin{equation}
v=\dfrac{\textnormal{Number of susceptible plants that become removed at equilibrium}}{\mathrm{\textrm{time to reach equilibrium}}}.\label{eq:2.16}
\end{equation}
}

\textcolor{black}{Finally, for the sake of simplicity, in this work
we suppose that the secondary crop of the intercropped systems is
not susceptible to the disease and, consequently, its plants can not
become infected (i.e. $p_{i}=0$ for every plant $i$ that belongs
to the secondary crop). However, note that the presence of a secondary
crop may modify the interactions between the plants of the main cultivar
(i.e. $d_{ij}$ and $\tilde{A}_{ij}$) and, consequently, their respective
probabilities $q_{i}(t-\tau)$. In the next section we define the
intercropping arrangements used in this work. Besides, the secondary
crop can be used to implement ``trap crops'', which may alter mobility
of an aphid, i.e. $\tilde{A}_{ij}$ (see subsection \ref{subsec:Implementation-of-the}).}

\subsection{Computational arrangements}

\subsubsection{\textcolor{black}{Intercropping arrangements.}}

\begin{figure}[H]
\begin{centering}
\includegraphics[width=0.9\textwidth]{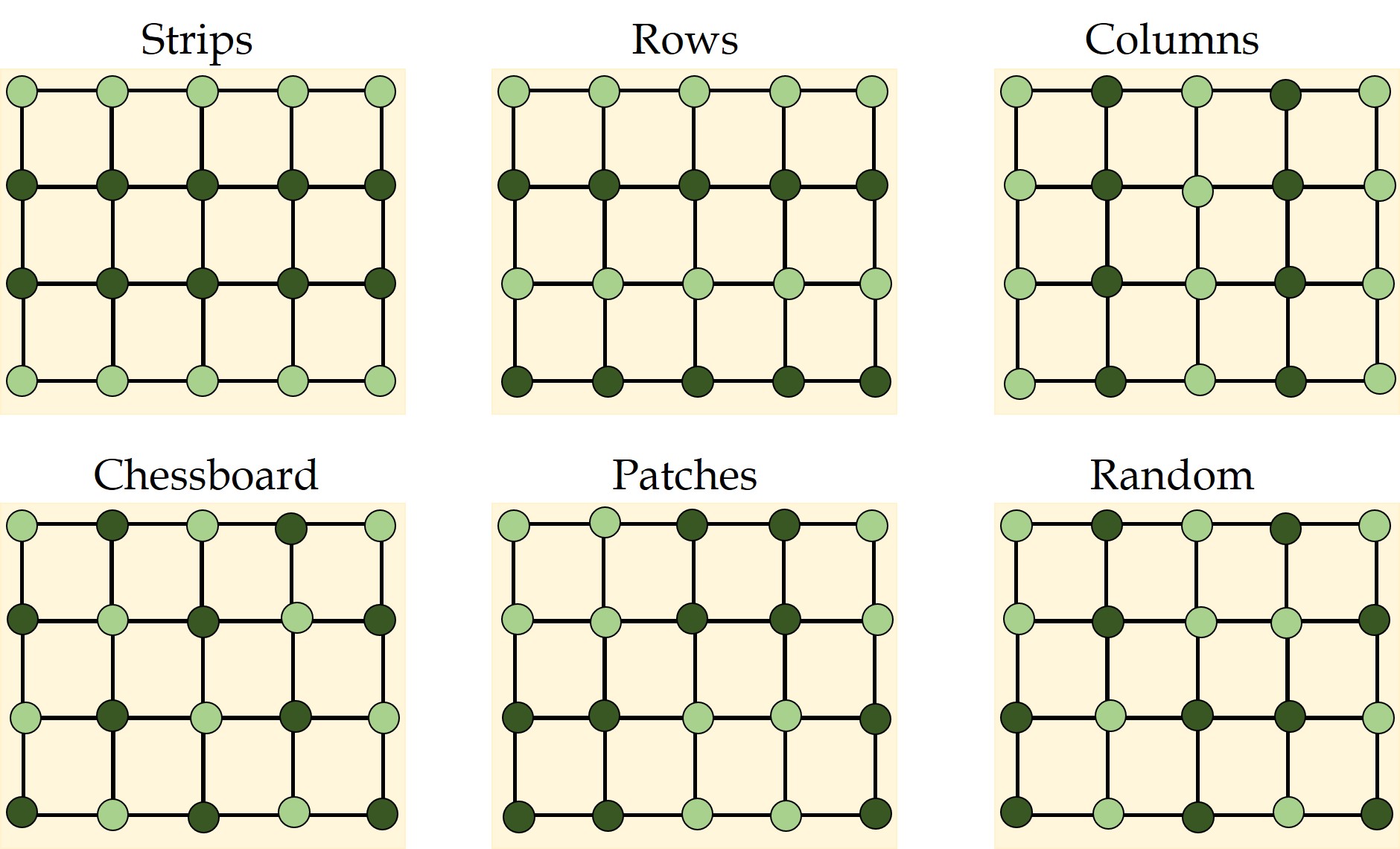}
\par\end{centering}
\caption{\textbf{Intercrop arrangements.} Different organizations of intercrops
between two species studied in this work with $r=\varDelta$ (see
Networks construction). Light green nodes represent the main crop
and the dark green nodes represent the secondary crop, which is considered
to be not susceptible to the disease spreading on the field. In the
case of trap crop strategies the dark green nodes represent the plants
with semiochemical activity to trap the pest to be controlled. The
square lattices connecting the nodes correspond to the interconnection
networks considered here.}

\label{intercroppings}
\end{figure}

\textcolor{black}{The intercropped systems considered here and shown
in Fig. \ref{intercroppings} are: the }\textbf{\textit{\textcolor{black}{strip}}}\textcolor{black}{{}
intercropping in which strips of the main cultivar are inserted between
strips of the secondary crop; }\textbf{\textit{\textcolor{black}{row}}}\textcolor{black}{{}
intercropping in which rows of the main and secondary crops are alternated
one-by-one; }\textbf{\textit{\textcolor{black}{column}}}\textcolor{black}{{}
intercropping, the same as before but by columns instead of by rows;
}\textbf{\textit{\textcolor{black}{chessboard}}}\textcolor{black}{{}
intercropping in which a plant of the main crop is inserted in the
rows and columns between every two susceptible plants; }\textbf{\textit{\textcolor{black}{patches}}}\textcolor{black}{{}
intercropping in which squared patches of the main crop are alternated
with squared patches (of the same size) of the secondary crop; }\textbf{\textit{\textcolor{black}{random}}}\textcolor{black}{{}
intercropping in which plants of the secondary crop are randomly inserted
among those of the main crop. The first two intercropping arrangements\textemdash strips
\cite{trap_crop_5,strips_1} and rows \cite{rows_1,rows_2,rows_3}\textemdash are
frequently used in experimental designs and field applications. It
is important to remark that in all cases we have considered exactly
the same amount of plants of the main crop such that the results obtained
here are not due to size effects. }

\subsubsection{\textcolor{black}{Networks construction.}\textcolor{red}{{} }}

\begin{figure}[H]
\begin{centering}
\includegraphics[width=0.45\textwidth]{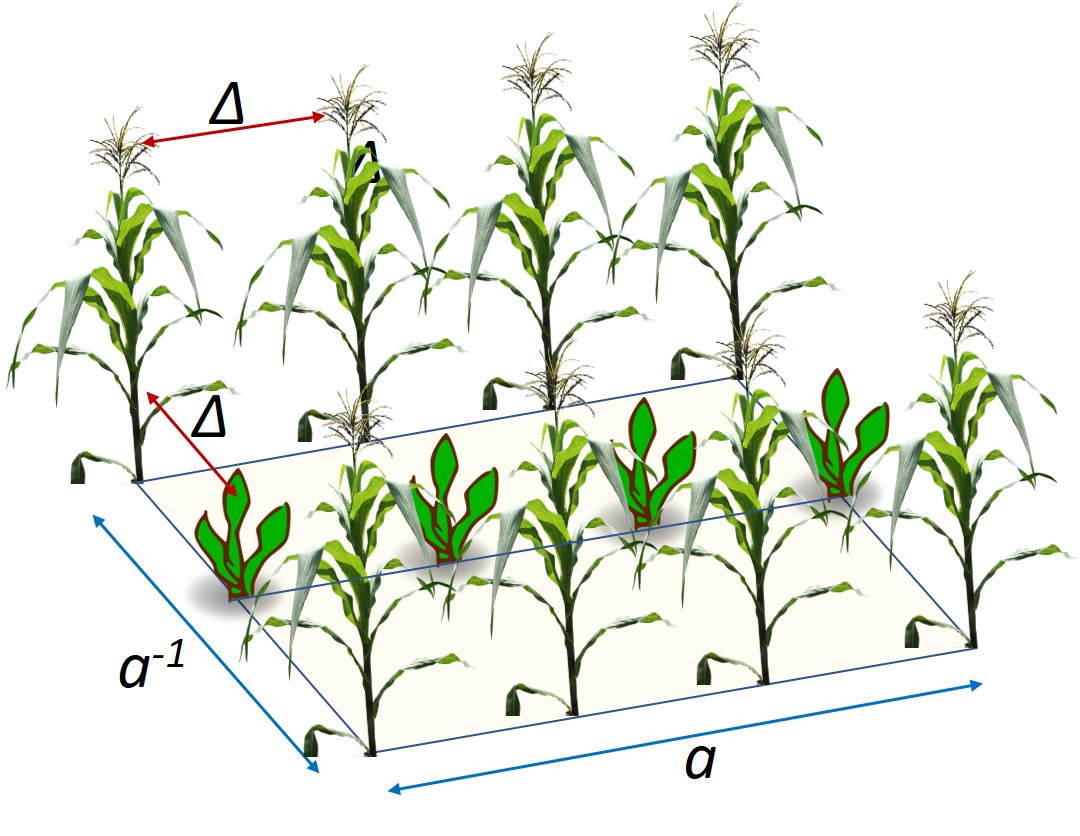}
\par\end{centering}
\caption{Schematic representation of the intercropping of two species in a
rectangular plot of unit area and largest edge length $a$. The separation
between plants is given by $\varDelta$. }

\label{field arrangement}
\end{figure}

Our arrangements consist of rectangular plots of lengths $x=a$ and
$y=a^{-1}.$ These plots guarantee that all simulations are carried
out on fields of equal area. \textcolor{black}{The rectangular plots
have been shown\textendash both theoretically and experimentally\textendash to
delay more the propagation of epidemics than square plots with the
same area and density of plants \cite{rectangular plots}. }We consider
the distribution of a major crop intercropped with a secondary crop,
which may or may not be a trap crop. In the intercropped field we
maintain a separation between plants equal to $\varDelta$ (see Fig.
\ref{field arrangement}). In this case the plant-to-plant connectivity,
based on their separation, is represented by a squared partition of
the plot. We simply normalize all the distances by dividing them by
$\varDelta$. Then, two plants which are nearest neighbors are one
step apart, a second nearest neighbor is two steps apart and so forth.
In general, every plot consists of 20 rows and 50 columns. There is
a plant at every intersection for a total of 1000 plants. As we have
a unit rectangle with $\ensuremath{a=1.6059}$, the value of $\varDelta$
is 0.033, and we use a connection radius $\ensuremath{r=\Delta},$
such that the plants are adjacent (connected in the network) only
to those immediately to the left, right, up and down. In the case
of the intercropped systems we always replaced 500 plants of the main
crop by the same quantity of plants of the secondary crop. In the
Supplementary Note 3 we analyze the case in which the separation between
rows and columns in the plot are smaller than $\varDelta$, which
is equivalent to consider the radius of primary movement of the aphid
equal to $r=\sqrt{2}\Delta$.

\subsubsection{\textcolor{black}{Implementation of the ``trap crops''.}\textcolor{red}{{}
\label{subsec:Implementation-of-the}}}

\begin{figure}[H]
\begin{centering}
\includegraphics[width=0.75\textwidth]{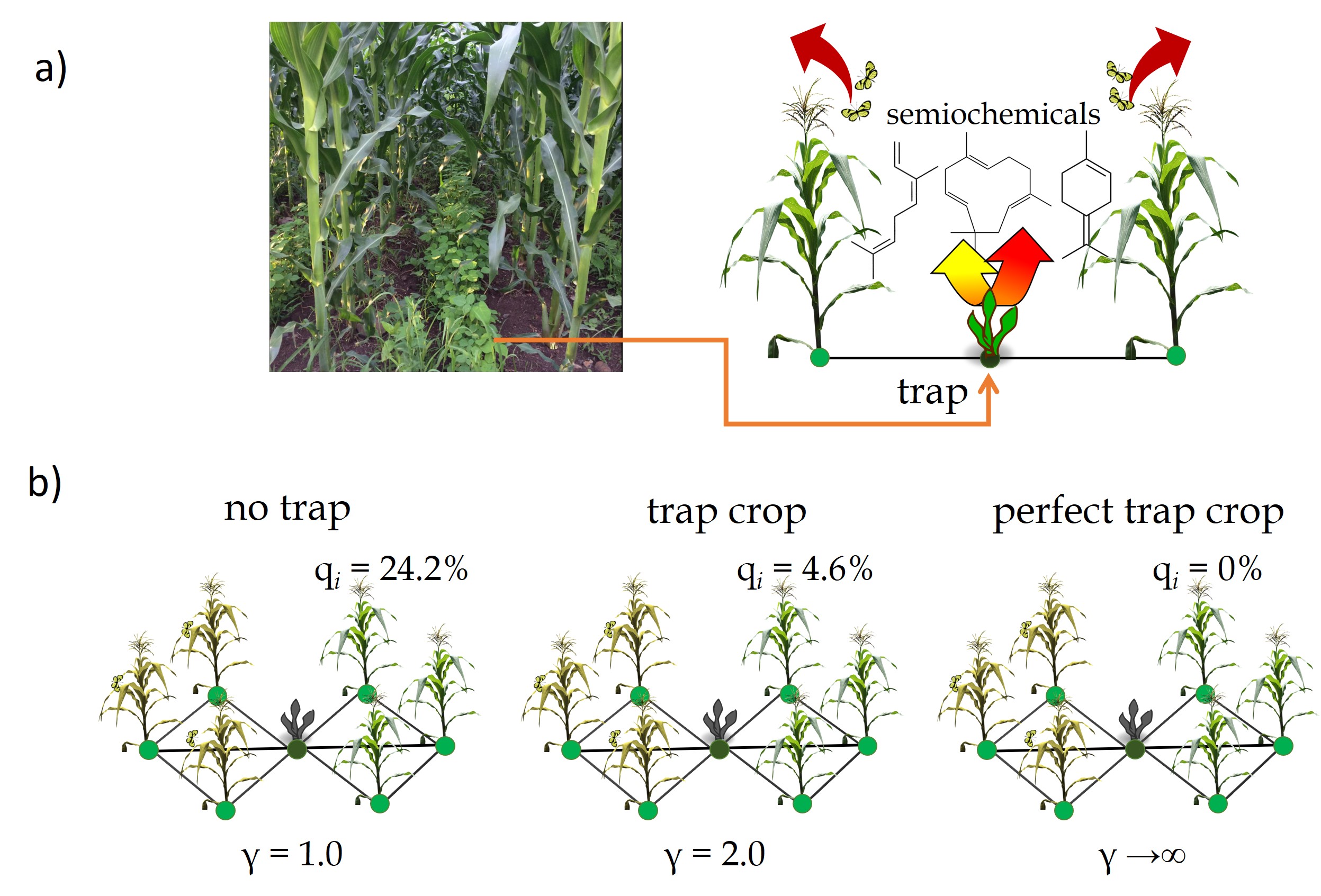}
\par\end{centering}
\caption{(a) Intercropping with 'push' (or 'push-pull') strategies where semiochemicals
\cite{semio_1,semio_2} are released from trap crops (the photograph
is courtesy of Rachel Monger (Immanuel International)). (b) Effects
of the strength of the trap crop $\gamma$\textcolor{black}{{} (dark
plants) on th}e probability of plants $i$ to get infected $q_{i}$
(see text for explanations) once the plants on the left of the Fig.
are infected. }

\label{traps examples}
\end{figure}

\textcolor{black}{Although trap crops can be formed either by ``push''
crops or by the combination of ``push-pull'' crops \cite{trap_crop_1,trap_crop_2,trap_crop_3,trap_crop_4},
here }for the modeling purpose we combine all the trap crop effects
into a single one. Basically we consider that trap crop diminishes
or completely avoids the propagation of a pest in a path beyond the
place in which the trap is located. Consequently, if there are more
than one trap in the path between two susceptible plants we only consider
the effect of one of them. An additive or multiplicative effect of
the traps can be easily implemented using the current mathematical
framework (see further), but it is not done here for the sake of simplicity.
In this case the secondary crop is located between the paths connecting
the infected and the susceptible plants. Mathematically, let us consider
two plants $i_{1}$ and $i_{d+1}$, and the \textit{shortest-path}
$i_{1}$, $i_{2}$, $\cdots$, $i_{d}$, $i_{d+1}$ of length $d$
between them. To model trap crops, we modify the strength of the long-range
mobility of the aphid between $i_{1}$ and $i_{d+1}$ as follows:

\begin{equation}
\left(\tilde{\mathbf{A}}\right)_{i_{1},i_{d+1}}=\left\{ \begin{array}{r}
d^{-\gamma s}\\
d^{-s}
\end{array}\right.\begin{array}{l}
\textnormal{if there is at least one trap crop between \ensuremath{i_{i}} and \ensuremath{i_{d+1}}}\\
\textnormal{otherwise}
\end{array},\label{adj_transf_1}
\end{equation}

\noindent where $\gamma\geq1$ is the \textit{\textcolor{black}{trap
strength}}. When $\gamma=1$, there is no trap crop as we recover
the original equation for the epidemic dynamics with long-range movements.
On the other hand, when $1<\gamma<\infty$, movement of the aphid
is reduced beyond the point in which the trap is located. For instance,
when the trap crop is very effective, i.e., $\gamma\rightarrow\infty$,
the movement of the aphid from $i_{1}$ to $i_{d+1}$ is completely
interrupted, which means that the trap is \textit{perfect}. In the
Fig. \ref{traps examples}(b) we illustrate the effects of a secondary
crop in which we obtained the probability $q_{i}$ that the plants
in the right part are infected once the three plants on the left side
are infected by the pest. \textcolor{black}{To do so, we suppose that
$p_{i}=0$ for $i$ spanning over the secondary crop and the plants
of the main cultivar that are in the right side of each arrangement,
and $p_{i}=1$ otherwise. According to Fig. \ref{traps examples},
the shortest path distance between the infected and the susceptible
plants is always $d=2$, and there is a secondary crop plant between
them. Under those conditions, Eq. (\ref{infect}) reduces to $q_{i}=1-\left(1-\beta2^{-s\gamma}\right)^{3}$
for each plant in the right side. Supposing $\tau=0$ (no delay),
$s=2.5$ (aphid with large mobility) and $\beta=0.5$, when $\gamma=1$
(no trap), the infectability of the susceptible plants is 24.2\%,
which represents the effects of an intercropped secondary species.
However, when the strength of the} trap is $\gamma=2$, the probability
that the susceptible plants are infected drops to less than 5\%. This
probability is reduced to zero as $\gamma$ is subsequently increased.

\subsubsection{Simulations. }

\textcolor{black}{Using the Markovian formalism, i.e. Eqs. (\ref{sir1})-(\ref{infect}),
we perform 100 random realizations for each field arrangement, secondary
crop (with or without trap) and aphid mobility (fast and slow). In
each independent realization, the propagation is initialized by infecting
randomly a single susceptible plant on the border of the field. Following
\cite{LRI_plants}, we set here $\mu=0.5$, since we are not trying
to characterize any particular disease. For $\mu=1$, for instance,
the recovery is too fast to see the spatial propagation and, conversely,
in the case $\mu=0$ the dynamics would be an SI dynamics. We decided
to lie between these two limiting cases.}

\textcolor{black}{For the dynamics of the disease we calculate the
total amount of Markovian time $t^{*}$ in which the probability of
being susceptible is larger than the probability of being removed
(i.e, $1-\varrho_{i}-p_{i}>\varrho_{i}$), for each plant of the main
cultivar. To estimate the epidemic thresholds for a given value of
$\gamma$, we calculate the average stationary fraction of removed
plants (over 100 independent realizations), $R(\beta,\mu)$, for 50
logarithmically spaced values of $\beta$, between 0.02 and 1.0, when
$\mu=0.5$, where $R(\beta,\mu)=\left(1/N_{i}\right)\lim\,_{t\rightarrow\infty}\sum_{i}\varrho_{i}(t)$,
$i$ spans over the plants of the main cultivar and $N_{i}$ is the
total amount of plants of that crop. Then, using a linear interpolation,
we find the epidemic threshold $\tau^{E}$ in each field. We recall
that the epidemic threshold is the smallest value of $\beta/\mu$
for each arrangement that satisfies the condition that $R(\beta,\mu)>0$.
Visualization of results in the form of rain clouds were performed
using Matlab\textsuperscript{\textregistered} codes available from Allen et al. \cite{visualization}}

\section{\textcolor{black}{Results and discussion}}

\subsection{\textcolor{black}{Influence of time-delay}}

\begin{figure}[H]
\subfloat[]{\textcolor{black}{\includegraphics[width=0.45\textwidth]{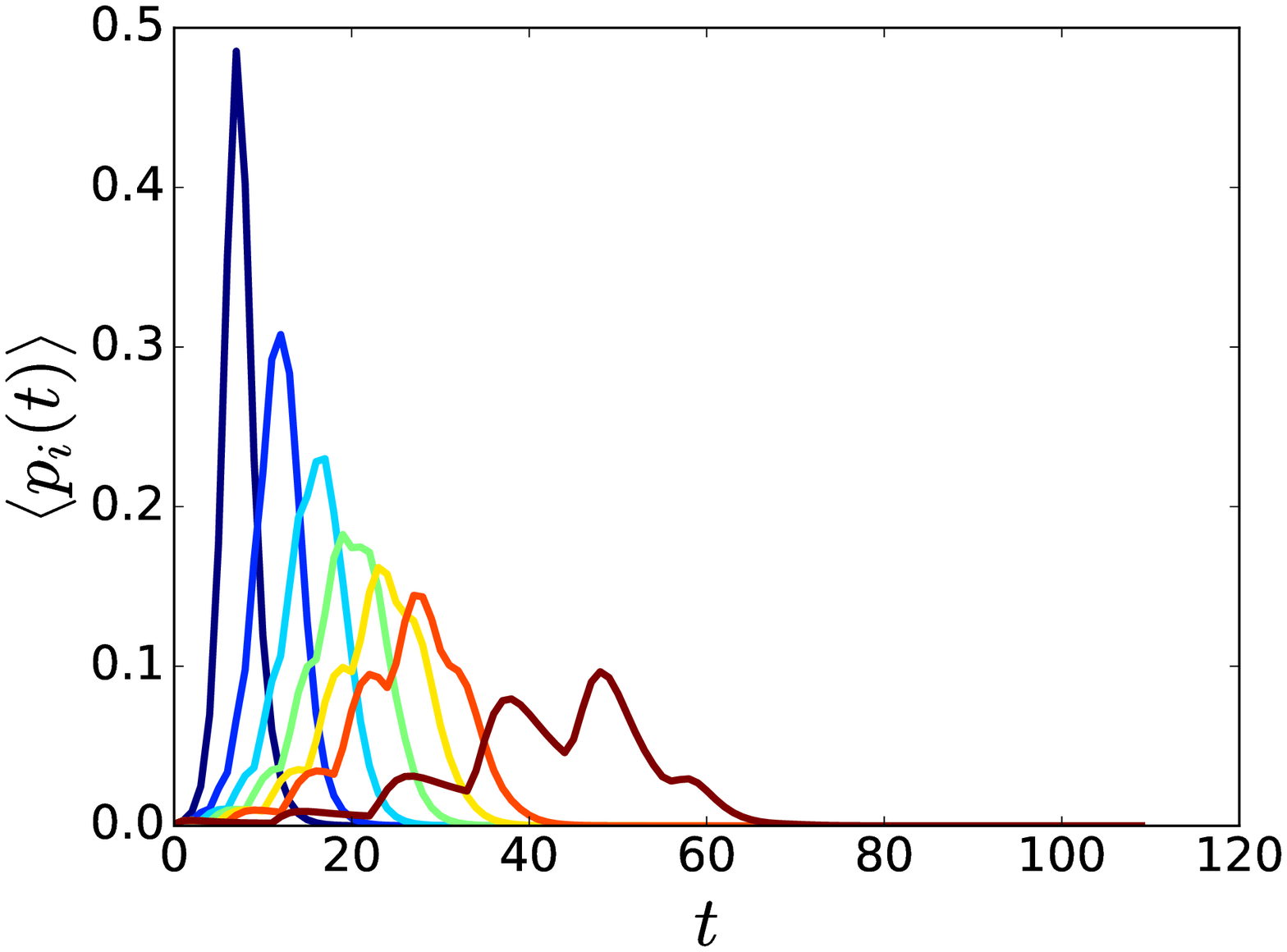}}

}\subfloat[]{\textcolor{black}{\includegraphics[width=0.45\textwidth]{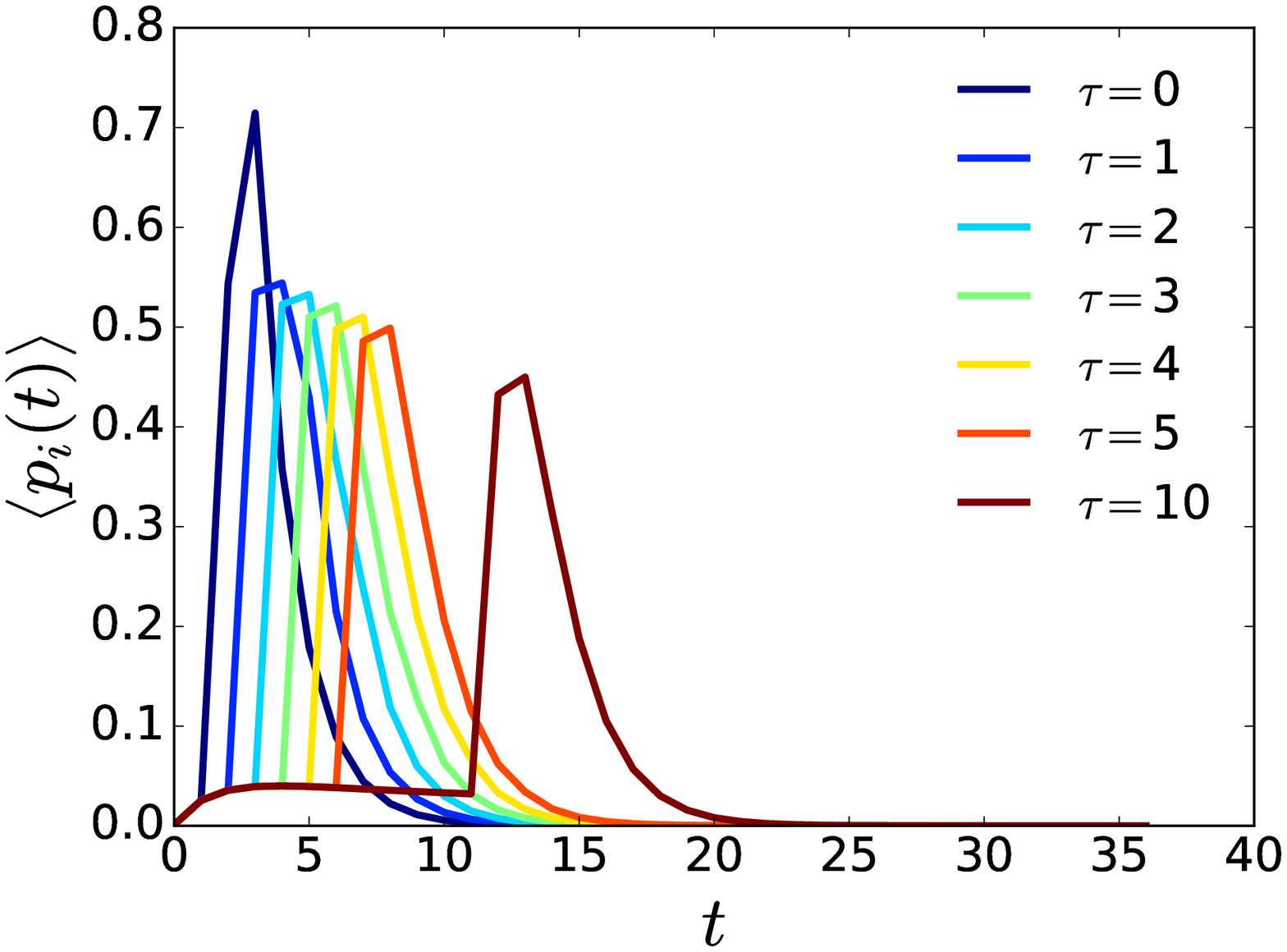}}

}

\caption{Evolution of the number of infected plants in a square plot with the
variation of the time delays $\tau$. The modeling is performed with
$\beta=0.5$, $\mu=0.5$, $r=\varDelta$ and $s=2.5$ (a) and $s=1.0$
(b).}

\label{time delay}
\end{figure}

\textcolor{black}{According to the results previously reported by
Tchuenche and Nwagwo \cite{Cooke_4}, the effects of the time delay
$\tau$ are mainly observed at the initial times of the propagation
dynamics and are focused on the population of susceptible plants.
For relatively large time the evolution of the SIR dynamics with and
without time delay are almost indistinguishable (see Fig. 2 in \cite{Cooke_4}).
For instance, in Fig. \ref{time delay} we can observe that the results
without time delay, i.e., $\tau=0$ are qualitatively similar to those
for $\tau=1$, with almost the same epidemic threshold and very similar
shape of the propagation curves. We explore here the effects of $\tau$
on the epidemic dynamics when the vector mobility is incorporated
into the model. Using the Markovian formulation described previously
in Eqs. (\ref{sir1})-(\ref{infect}), we obtained the evolution of
the infected population of plants in crop field consisting of a square
lattice as described before for two different values of the aphid
mobility $s$ in the Mellin transformed Markovian SIR equations. The
results are illustrated in Fig. \ref{time delay} were we have used
$\beta=0.5$, $\mu=0.5$, $r=\varDelta$ and $s=2.5$ (a) and $s=1.0$
(b). It can be seen that the inclusion of a time delay in the model
makes that the peak in the number of infected plants is displaced
to longer times. For large aphid mobility ($s=1.0$) it is observed
that the shapes of the peaks of infection are very similar to each
other for different values of the time delay $0\leq\tau\leq10$. When
the mobility of the aphids is relatively low ($s=2.5$) the rate of
propagation of the infection changes significantly for different values
of $\tau$, particularly for very large time delays. For instance,
the values of the rate of propagation for a given time-delay, $v\left(\tau\right)$,
obtained from Eq. (\ref{eq:2.16}), are as follow: $v\left(0\right)=32.25$,
$v\left(1\right)=26.32$, $v\left(2\right)=22.22$, $v\left(3\right)=18.87$,
$v\left(4\right)=16.67$, $v\left(5\right)=14.70$, $v\left(10\right)=9.17$.
However, for the case of large aphid mobility these rates of propagation
are not changed significantly with the time delay: $v\left(0\right)=43.48$,
$v\left(1\right)=40.00$, $v\left(2\right)=38.46$, $v\left(3\right)=35.71$,
$v\left(4\right)=34.48$, $v\left(5\right)=33.33$, $v\left(10\right)=27.78$.
That is, for relatively low time delays the results in the disease
propagation on plants are very similar to those without time-delays.
Also, when the the aphid mobility is relatively large, the time delay
does not affect significantly the propagation rate of the disease.}

As a consequence of the previous analysis and for the sake of keeping
our model as simple as possible we are not considering explicitly
the time delay in the further calculations in this work. The biological
justification for this simplification is as follows. The interaction
of the virus and aphid is controlled by the following phases (see
Chapter 15 of \cite{Aphids book}): (i) acquisition, where the aphid
takes up virions from an infected plant, (ii) retention, where the
aphid carries the virions at specific sites, (iii) latency, which
refers to the inability of an aphid to inoculate immediately a virus
following acquisition, and (iv) inoculation, which is the release
of retained virions into the tissues of a susceptible plant. There
are three types of transmission of a virus to a plant (see Chapter
15 of \cite{Aphids book}). In the \textit{non-persistent} (NP) transmission,
the acquisition and inoculation are very fast and requires only a
very brief stylet penetration, which delays less than one minute.
In this case there is no latency period and the whole cycle of transmission
can be completed within a few minutes. In the \textit{semi-persistent}
(SP) transmission, the acquisition and inoculation requires periods
of about 15 minutes. In this case there is no latency periods either
and the aphids retain the ability to inoculate for periods of up to
2 days following acquisition. Finally, in the \textit{persistent}
(P) transmission the virus acquisition requires period between hours
to days, there is a latency period and the retention is for days to
weeks. From the about 270 viruses transmitted by aphids more than
200 are transmitted by NP transmission (see Chapter 15 of \cite{Aphids book}).
The results to be considered here using a SIR model without time delays
is then equivalent to model the aphid-borne transmission of viruses
to plants using either NP or SP transmission.

\subsection{Impact of intercrop arrangements on virus propagation.}

\begin{figure}[H]
\begin{centering}
\subfloat[]{\begin{centering}
\includegraphics[width=0.47\textwidth]{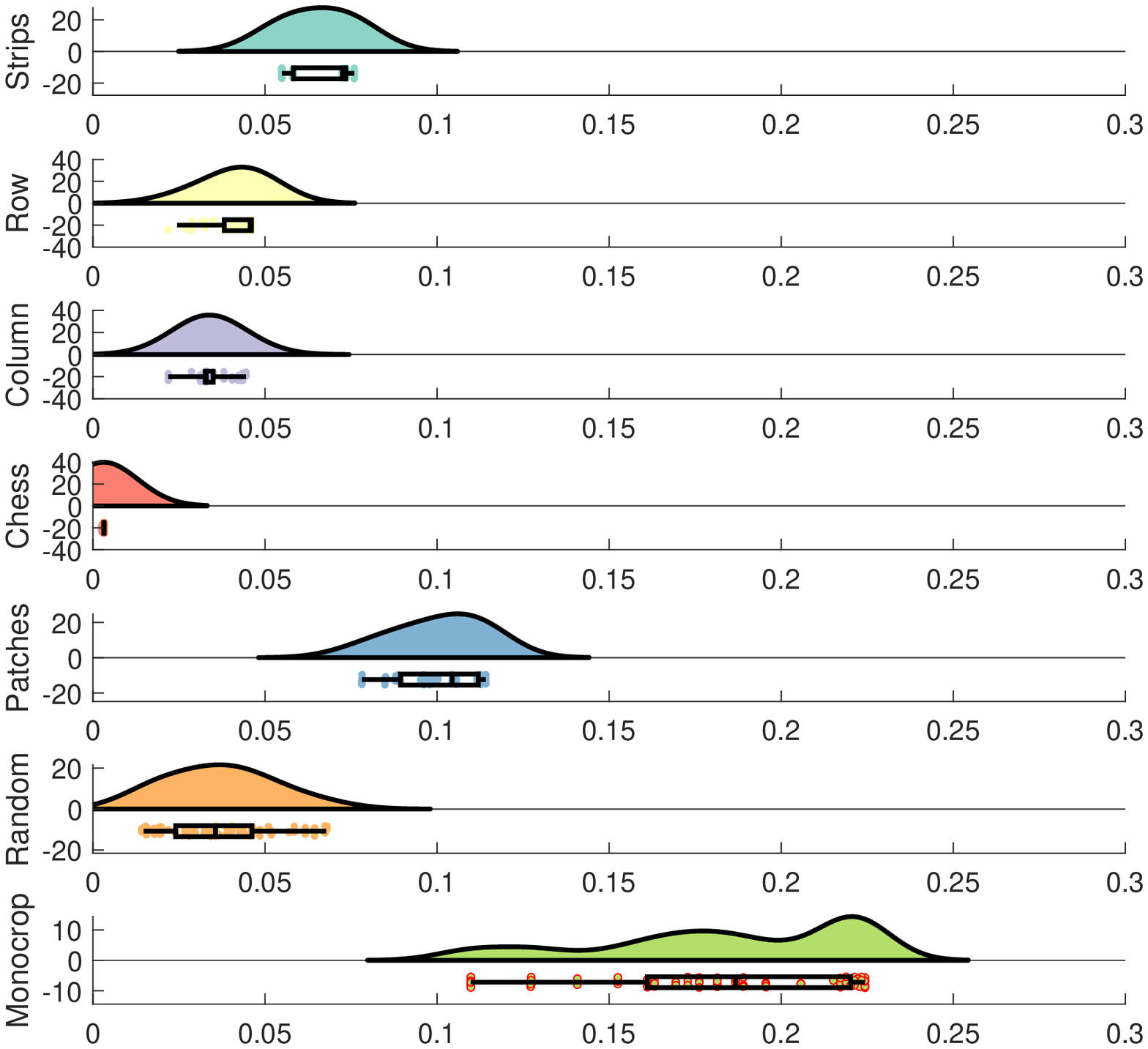}
\par\end{centering}

}\subfloat[]{\begin{centering}
\includegraphics[width=0.47\textwidth]{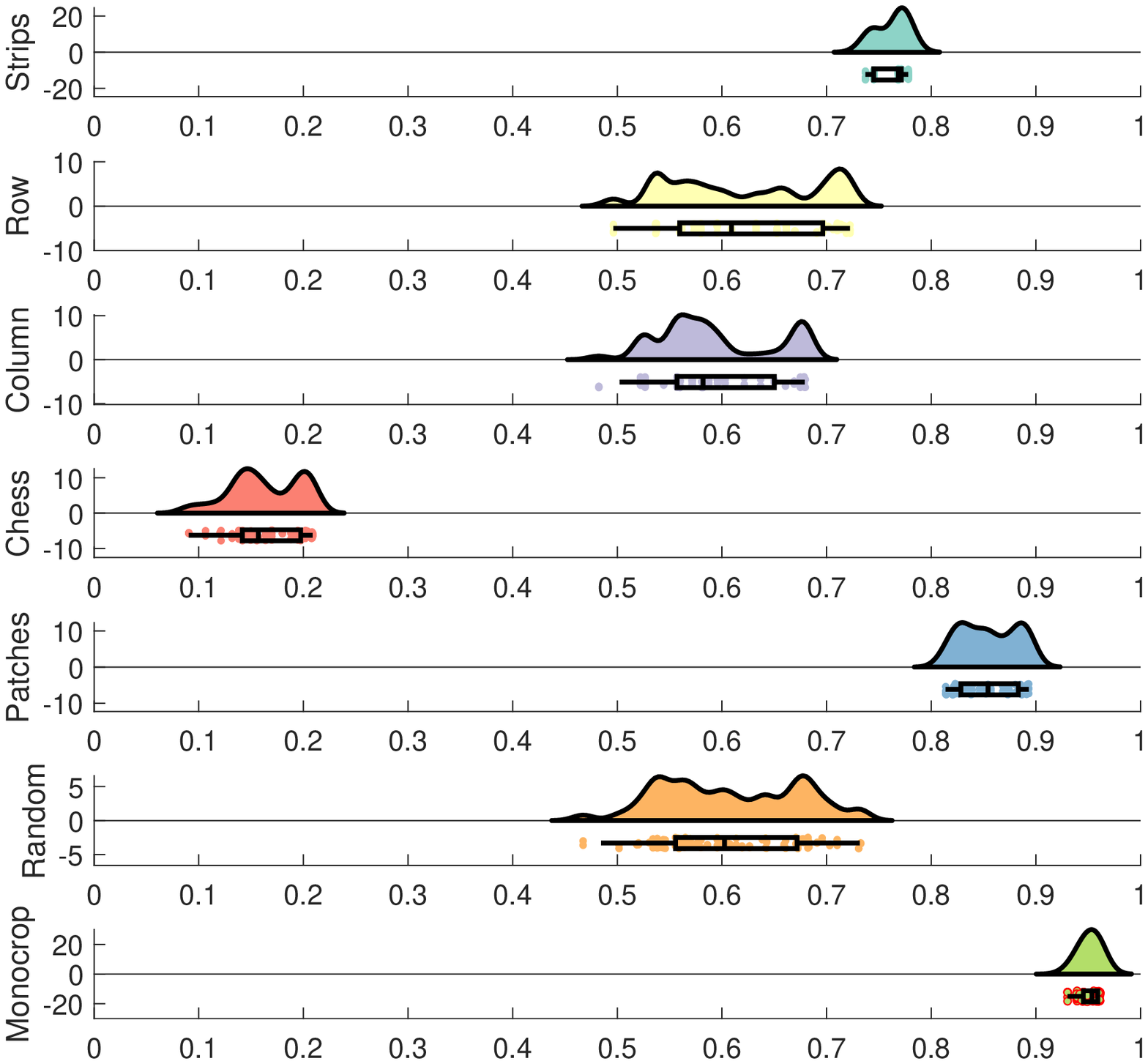}
\par\end{centering}
}
\par\end{centering}
\begin{centering}
\subfloat[]{\begin{centering}
\includegraphics[width=0.47\textwidth]{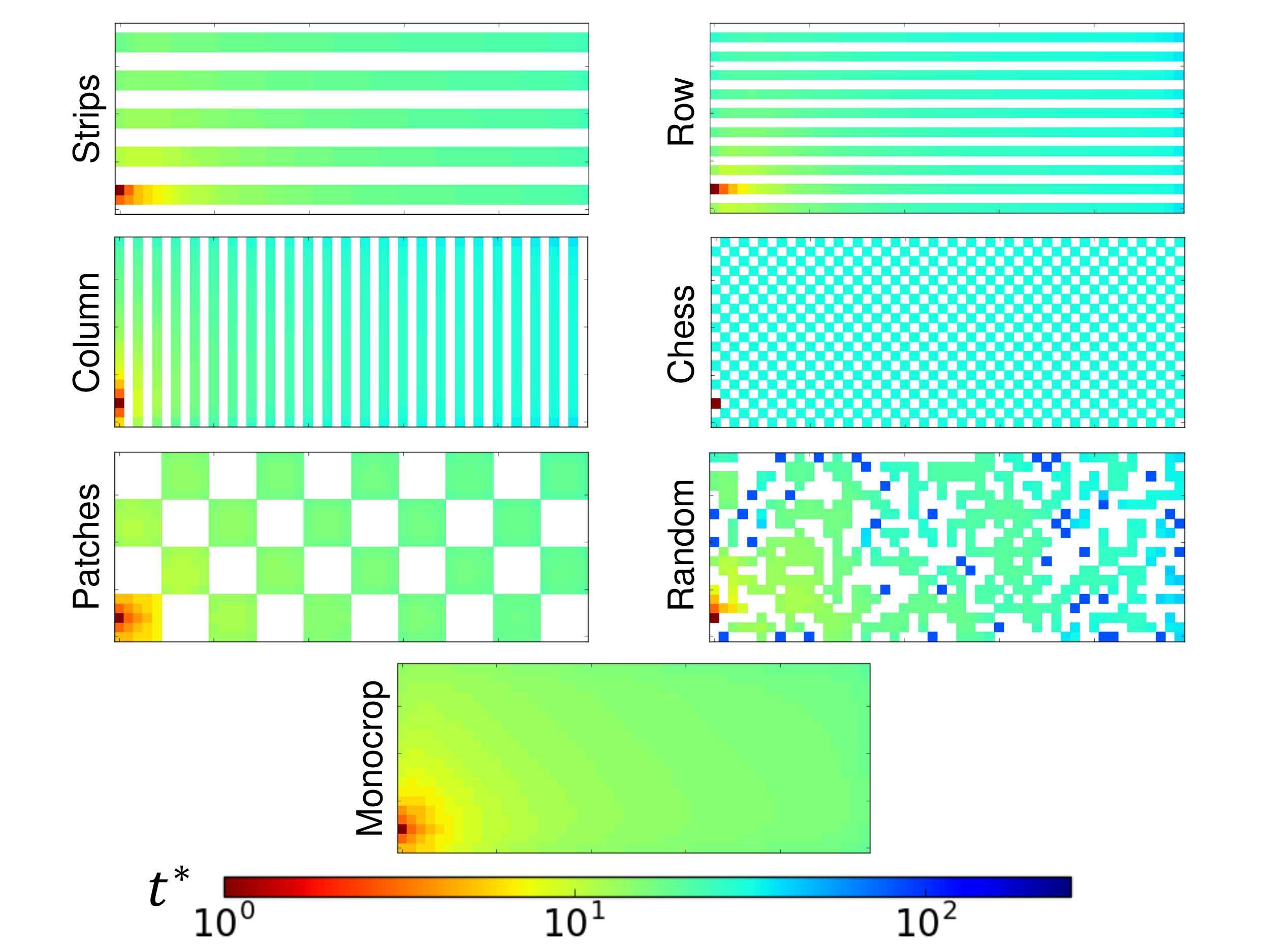}
\par\end{centering}
}\subfloat[]{\begin{centering}
\includegraphics[width=0.47\textwidth]{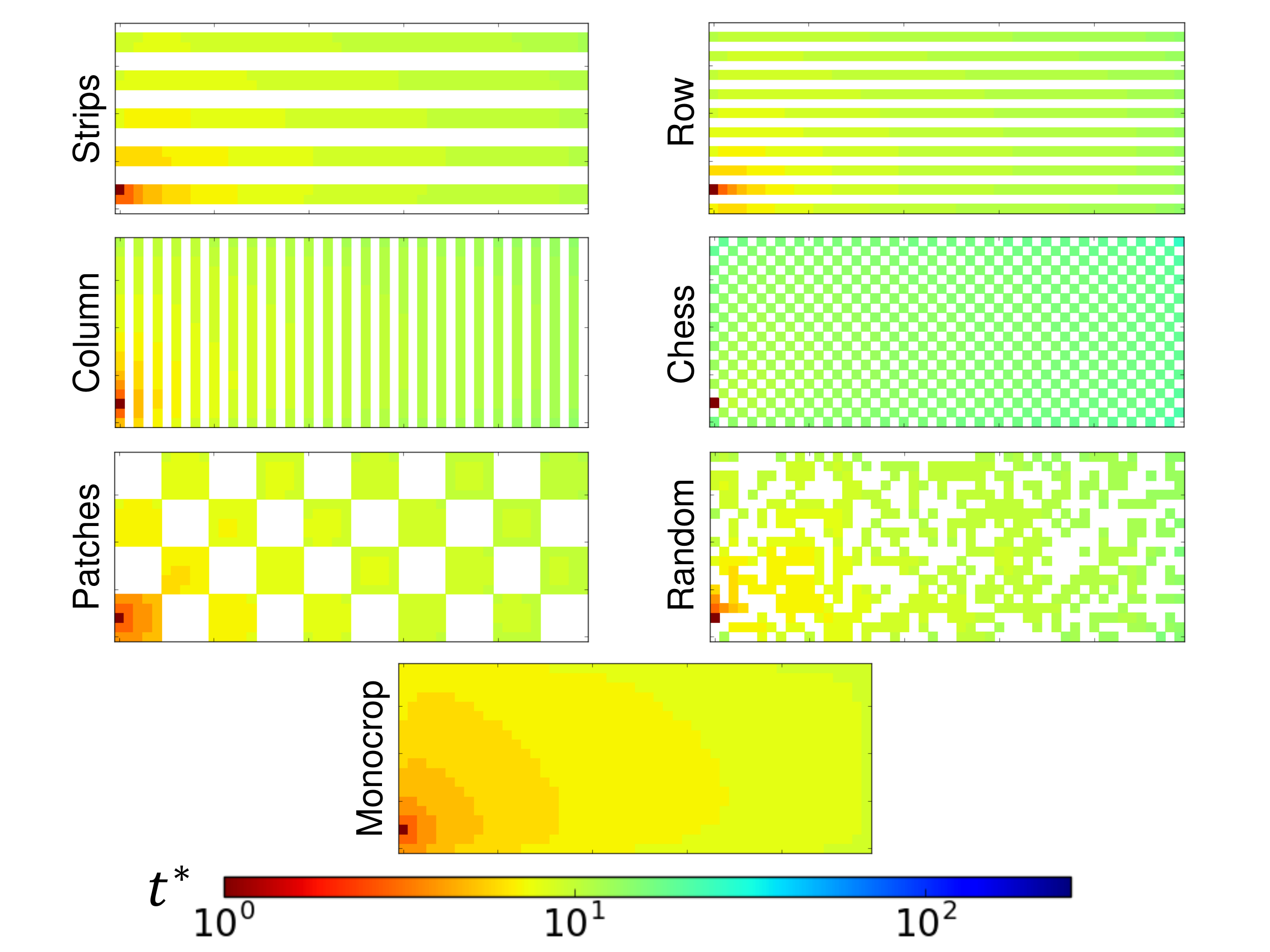}
\par\end{centering}
}
\par\end{centering}
\caption{\textbf{Aphid-borne virus propagation on intercropped fields without
traps.} Results of the simulations for a SIR epidemics at $t=10$
with $r=\varDelta,$ $\beta=0.5,$ $\mu=0.5$ for different intercropping
strategies without trap crops, i.e., the strength of the trap crop
is $\gamma=1.0$. Raincloud plots of the proportion of dead plants
for a viral infection propagated by aphids: (a) Aphid with reduced
mobility ($s=4.0)$ and (b) with larger mobility ($s=2.5$). The clouds
show the kernel distribution of the proportion of dead plants for
different realizations of the epidemics. Below, the raw data is plotted
(the rain) together with their corresponding box and whisker plots.
Illustration of the evolution of infection across fields with different
intercropping systems: (c) Aphid with reduced mobility ($s=4.0)$
and (d) with larger mobility ($s=2.5$). In both panels the time $t^{*}$
is given in a color scale (see text), and the propagation is initialized
by infecting the plant on the bottom-left corner of the plot.}

\label{No_traps}
\end{figure}

\textcolor{black}{In Fig. \ref{No_traps} we illustrate the results
of the simulations of the propagation of an aphid-borne virus in the
6 intercropped fields without traps $(\gamma=1.0$) studied here as
well as in the monocrop. In Fig. \ref{No_traps} (a) we show the results
for an aphid with relatively low mobility ($s=4.0$) and in Fig. \ref{No_traps}
(b) we give the same for a relatively high mobility aphid ($s=2.5$).
To compare the dynamics of the different arrangements, firstly, we
analyze their respective results before they reach equilibrium ($t=10$).
In the case in which $s=4.0$ it is clear that the disease is propagated
in a relatively slow fashion and for $t=10$ only 18.3\% of plants
are removed in the monocrop. As can be seen in this figure all intercrop
arrangements produce significant decrease in the number of removed
plants. The smallest decay in the number of removed plants is observed
for the patches configuration in which the percentage of removed plants
is 10.1\%, followed by the strips configuration with 6.6\%. On the
other hand, the most efficient arrangement is the chessboard one,
which reduces the number of removed plants practically to zero (only
0.3\% of removed plants). }

\textcolor{black}{In Fig. \ref{No_traps} (b) we illustrate the results
for the case in which the pest has a relatively large mobility. Here
the picture observed is significantly different from the one in the
previous case. First, the level of plants removed in the monocrop
is 95.1\%, indicating an almost complete destruction of the crop in
a relatively short time ($t=10$) when the pest is highly mobile.
The range of amelioration of the infection across the fields is here
very wide, ranging from the 10\% of decrease in removed plants observed
for the patches arrangement (85.5\% of removed plants) up to about
80\% of decrease obtained with the chessboard arrangement (16.3\%
of removed plants). Notice that the frequently used intercrop arrangement
of strips produces, together with that of patches, the smallest improvement
in the number of removed plants. Thus, although the results are quantitatively
very different for the cases of low and high mobility of the aphid,
they are qualitatively similar in identifying the worse arrangements
(patches and strips) as well as the best one (chessboard). In both
cases the order of effectivity in reducing the impact of an aphid-borne
virus propagation is: chessboard > columns > random > rows > strips
> patches.}

\textcolor{black}{In Fig. \ref{No_traps} (c) and (d) we illustrate
a snapshot of the aphid-borne propagation of a virus across the different
intercropping systems with $s=4.0$ and $s=2.5$, respectively. In
order to compare all the different arrangements we always start the
epidemic by infecting the same node, i.e., the one at the bottom-left
corner of the field. The colors in the plots represent the time $t^{*}$
in which the plant remains susceptible without becoming removed by
the vector-borne virus disease. That is, a low value of this time
indicates that the plant is removed relatively soon by the virus disease.
In order to interpret quantitatively the results in these plots we
use the rate $v$ of propagation of the aphid-borne virus previously
defined in Eq. (\ref{eq:2.16}). It can be seen that in the monocrop
the epidemic is propagated in a wave-like way, typical of diffusion
processes. The values of $v$ in the monocrop are 23.26 ($s=4.0$)
and 32.26 ($s=2.5$). That is, when the aphid has relatively low mobility
there is an average infection of 23.26 plants per unit time. This
rate is increased to 32.26 plants when the pest mobility is increased,
due to the fact that the aphids can now hop to wider regions of the
plots. Reminiscences of the wave-like kind of propagation of the vector-borne
virus are observed in all the intercrop arrangements studied. In the
intercropped systems (without trap crops, $\gamma=1.0$) the propagation
rates of the virus are: for $s=4.0$, chessboard (0.03) < random (5.46)
< columns (7.35) = rows (7.35) < strips (10.0) < patches (11.36);
for $s=2.5$, chessboard (9.62) < random (12.19) < columns (13.16)
= rows (13.16) < strips (14.70) < patches (15.62). In closing, the
chessboard arrangement is significantly better in reducing the propagation
of aphid-borne viruses in agricultural fields than the rest of the
arrangements when there are no trap crops in the intercrop. The random
arrangement also performs very well in terms of both the number of
plants removed by the infection and the rate of propagation of the
epidemic.}

\textcolor{black}{Finally, it is worth recalling that the prior results
depend on the radius for primary dispersal of the aphid. See Supplementary
Note 3 for the case when the separation between rows and columns is
smaller than here and the pest can hop not only across the rows and
columns, but also diagonally between rows, i.e., when the radius for
primary dispersal of the aphid is $r=\sqrt{2}\varDelta$ instead of
$r=\varDelta.$ When the pest mobility is relatively low ($s=4.0$),
the best arrangements are the rows and columns intercrops with about
10\% of affected plants vs. 78\% affected in the monocrop for $t=10$.
However, when the pest has high mobility ($s=2.5$), none of the intercropping
systems is able to stop the propagation of the pest across the field,
with percentages of affected plants similar to that in the monocrop
(98.6\%). An obvious measure to mitigate this problem is to increase
the separation of the rows and columns in the crop field, or even\textendash as
shown in the experiments by Khan et al. \cite{Khan_et_al}\textendash to
increase the separation between rows keeping a smaller separation
between columns.}

\subsection{Impact of intercrops with trap crop on aphid-borne virus propagation. }

\begin{figure}[H]
\begin{centering}
\subfloat[]{\begin{centering}
\includegraphics[width=0.47\textwidth]{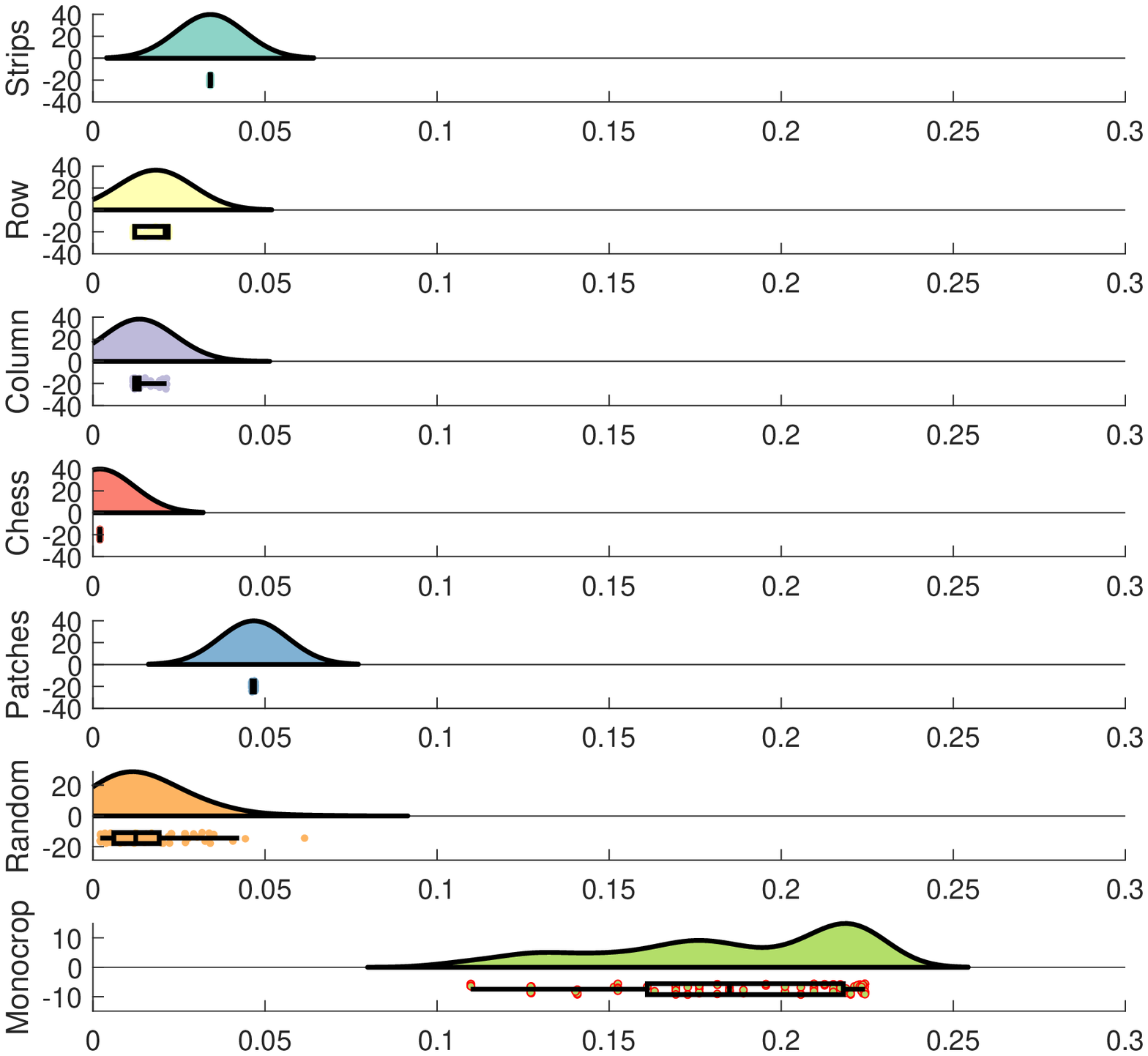}
\par\end{centering}
}\subfloat[]{\begin{centering}
\includegraphics[width=0.47\textwidth]{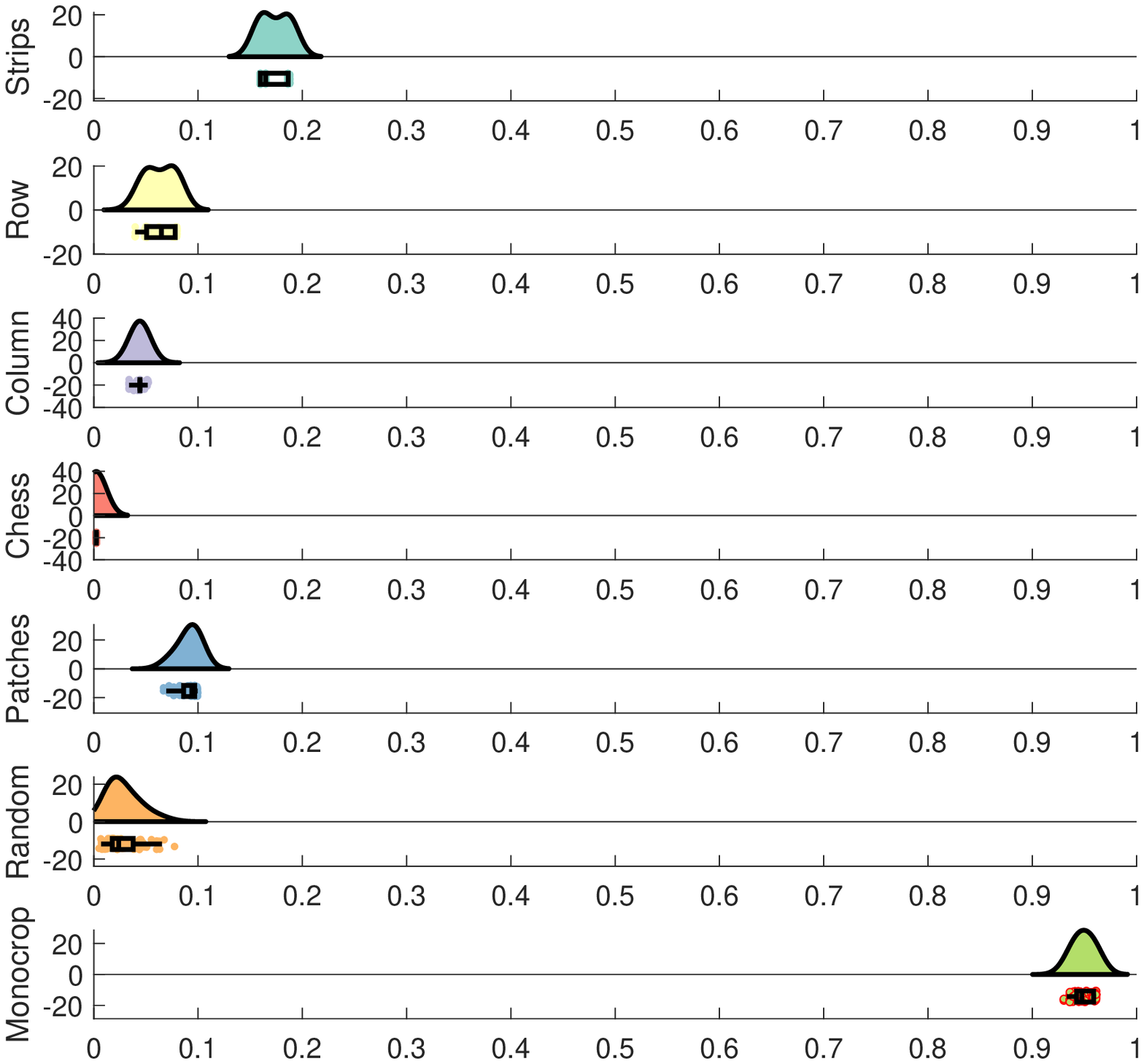}
\par\end{centering}
}
\par\end{centering}
\begin{centering}
\subfloat[]{\begin{centering}
\includegraphics[width=0.47\textwidth]{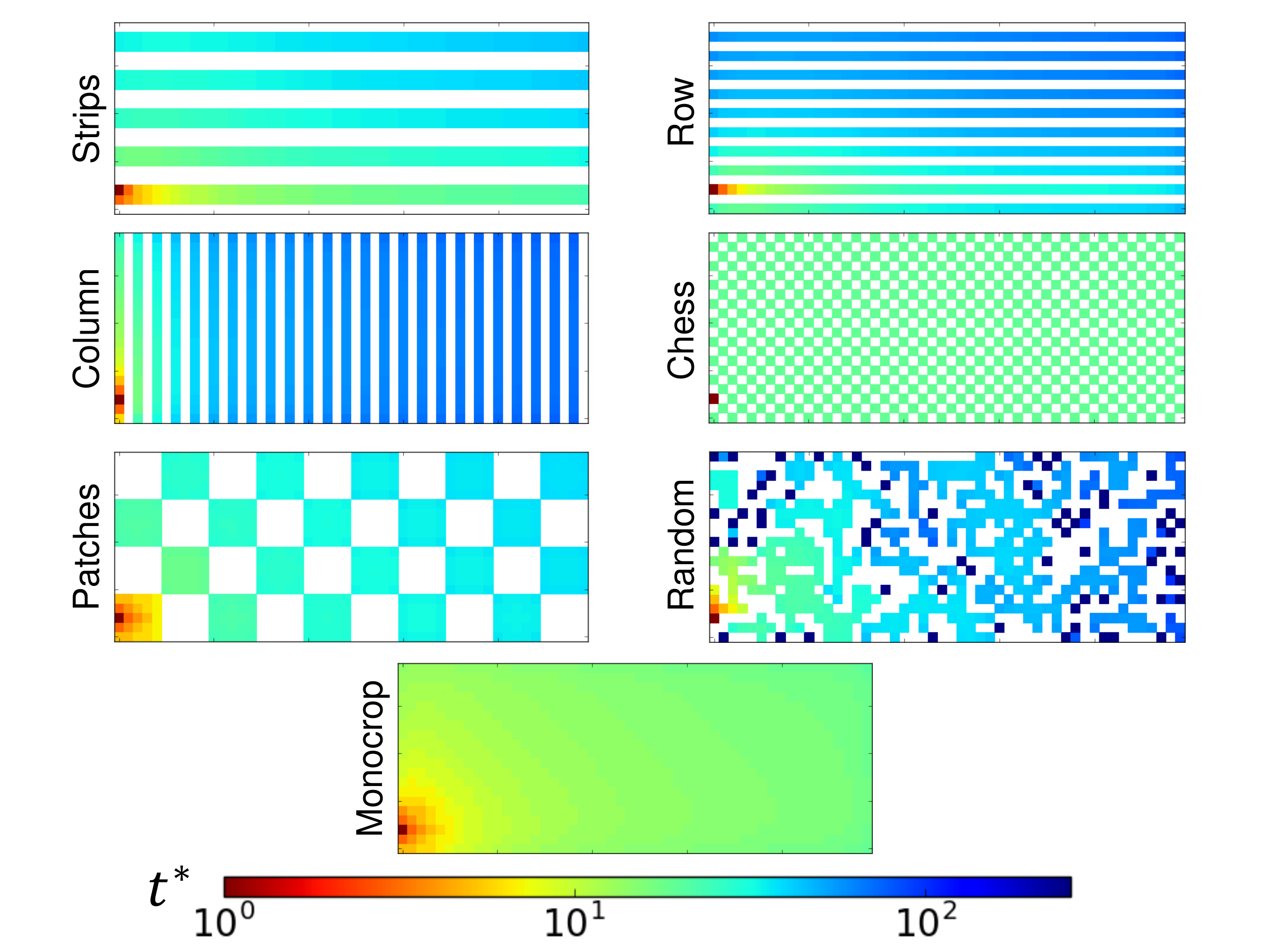}
\par\end{centering}
}\subfloat[]{\begin{centering}
\includegraphics[width=0.47\textwidth]{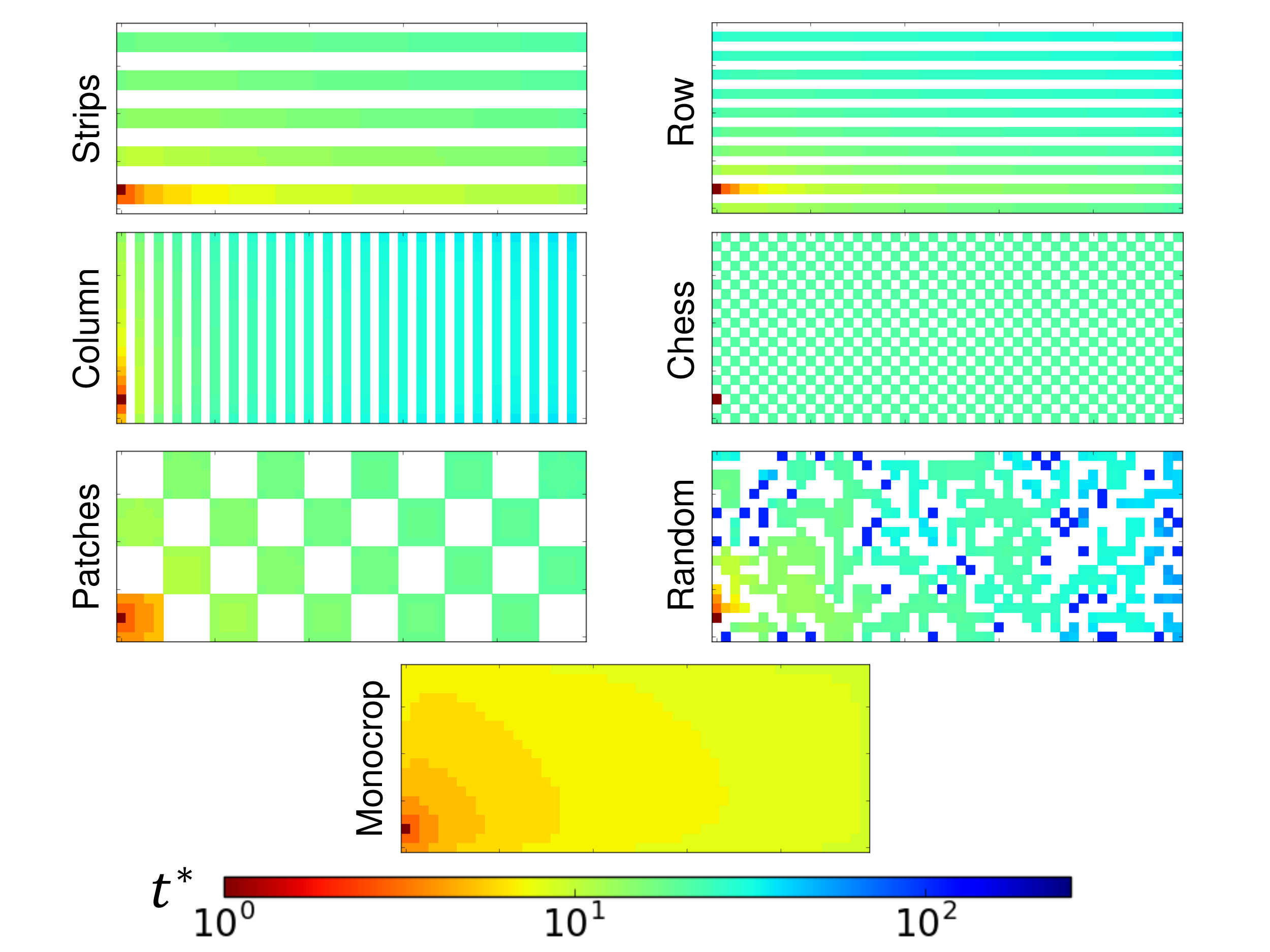}
\par\end{centering}
}
\par\end{centering}
\caption{\textbf{Aphid-borne virus propagation on intercropped fields with
trap crops.} Results of the simulations for a SIR epidemics at $t=10$
with $r=\varDelta,$ $\beta=0.5,$ $\mu=0.5$ for different intercropping
strategies with trap crops of strength $\gamma=2.0$. Raincloud plots
of the proportion of dead plants for a viral infection propagated
by aphids: (a) Aphid with reduced mobility ($s=4.0)$ and (b) with
larger mobility ($s=2.5$). The clouds show the kernel distribution
of the proportion of dead plants for different realizations of the
epidemics. Below, the raw data is plotted (the rain) together with
their corresponding box and whisker plots. Illustration of the evolution
of infection across fields with different intercropping systems: (c)
Aphid with reduced mobility ($s=4.0)$ and (d) with larger mobility
($s=2.5$). In both panels the time $t^{*}$ is given in a color scale
(see text), and the propagation is initialized by infecting the plant
on the bottom-left corner of the plot.}

\label{Traps}
\end{figure}

We now move to the analysis of the intercrop systems with trap crops.
To have an idea of the many systems in which the current results can
be applied the reader is referred to the Tables 1 and 2 in Hokkanen's
paper \cite{trap_crop_1}, where many examples of one main crop intercropped
with a trap crop are given. We consider here the existence of trap
crops which are not perfect, i.e., they allow certain propagation
of the aphid-borne viral infection (see Supplementary Note 4 for results
with a perfect trap). Thus, we use $\gamma=2.0$ and analyze the cases
of relatively low ($s=4.0$) and relatively large ($s=2.5$) aphid
mobility. In Fig. \ref{Traps} we illustrate the results of our simulations
for these systems using the different arrangements studied here. As
can be seen for the case of relatively low mobility ($s=4.0$) there
are significant reduction in the percentages of removed plants for
all intercrop systems. The percentages of removed plants for each
intercrop are: chessboard (0.2\%), columns (1.4\%), random (1.5\%),
rows (1.8\%), strips (3.4\%) and patches (4.7\%). We remind the reader
that the percentage of removed plants in the monocrop is 18.1\%. When
the pest has a relatively large mobility ($s=2.5$), 95.1\% of plants
are removed in the monocrop, while in each of the intercrops they
are: chessboard (0.2\%), random (2.8\%), columns (4.4\%), rows (6.3\%),
patches (9.1\%), and strips (17.4\%). Notice that here there are some
important changes in the order of the arrangements in terms of their
effectivity in reducing the propagation of the infection. When the
aphid is of high mobility the best arrangements are the chessboard
and the random one. The worse arrangement, and the only one having
more than 10\% of removed plants, is the strip one. Also notice that
the percentage of removed plants in the chessboard arrangement is
exactly the same for $s=2.5$ and $s=4.0$, indicating a high stability
in the efficiency of this arrangement. It is important to remark one
more time that these reductions in the number of removed plants are
the consequence of the different topological patterns emerging from
the intercrop arrangements. That is, these differences are not a dilution
effect due to the fact that the number of susceptible and immune plants
are kept the same in every arrangement.

We now analyze the rate of propagation of the aphid-borne virus across
the agricultural fields intercropped with a trap crop (see Fig. \ref{Traps}
(c) and (d)). The rate of propagation of the virus follow a different
order as for the case of intercrops without traps ($\gamma=1.0$).
That is, for $s=4.0$, we find: chessboard (0.05) < random (1.67)
< columns (4.59) < rows (4.67) < strips (7.04) < patches (7.94). For
$s=2.5$, chessboard (0.04) < random (4.18) < columns (7.58) < rows
(8.06) < strips (10.87) < patches (11.63). Here again there is a significantly
high improvement, in terms of diminishing the impact and the rate
of propagation of a virus across an agricultural field, when the chessboard
arrangement is used. See Supplementary Note 3 for the case when the
separation between rows and columns is smaller than here, i.e., when
the radius for primary dispersal of the aphid is $r=\sqrt{2}\varDelta$
instead of $r=\varDelta.$ These results agree with those previously
reported using a \textcolor{black}{different }stochastic simulation
model \cite{stochastic model}.

\subsection{Epidemic thresholds.}

\begin{figure}[H]
\begin{centering}
\includegraphics[width=0.9\textwidth]{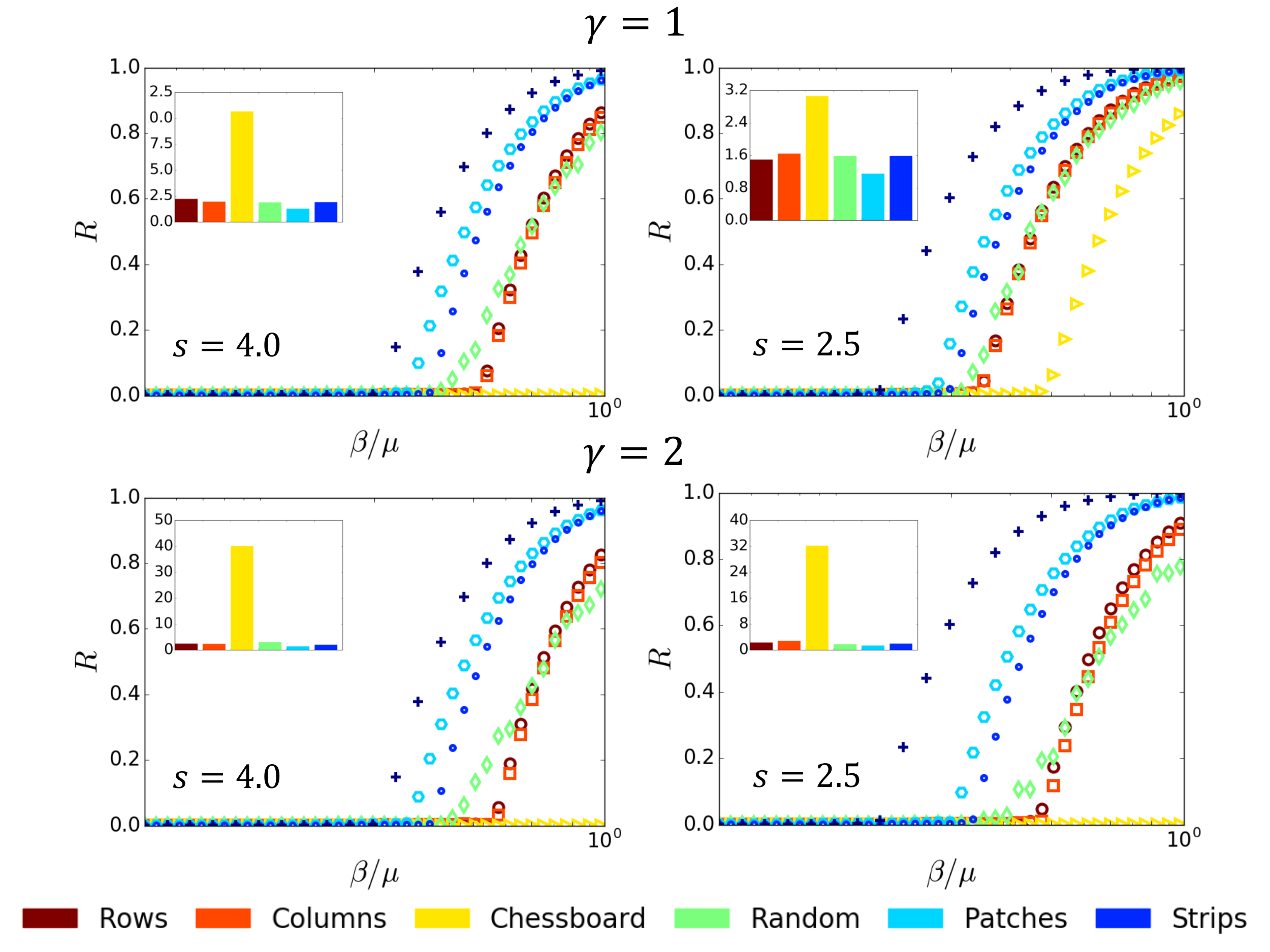}
\par\end{centering}
\caption{\textbf{Epidemic thresholds for the propagation of aphid-borne viruses
on intercropped fields. }Effects of the strength of the trap crop
$\gamma$ and of the pest mobility $s$ on the number of removed (dead)
plants during the propagation of an aphid-borne virus infection across
intercropped fields with $r=\varDelta$. The case for $\gamma=1$
corresponds to no trap crop. The symbols used for each intercrop are:
strips (small circles), rows (large circles), columns (squares), chessboard
(triangles), patches (hexagons), random (diamonds), and monocrop (crosses).
In the insets we illustrate the normalized epidemic thresholds for
each of the intercrop arrangements. The normalization is obtained
by dividing every epidemic threshold by that of the monocrop in the
corresponding system. The color code for the insets is given at the
bottom of the figure.}

\label{thresholds}
\end{figure}

\textcolor{black}{Finally we study }the ratio $\beta/\mu$, which
drives the spreading of the disease. Depending on the infectious power
of the aphid-borne virus there are two possible distinguishable phases
for a given strength of the trap crop, $\gamma$, and of the pest
mobility, $s$. The first one is an absorbing phase where the spreading
of the virus is not efficient enough to reach a large fraction of
the system and the propagation is absorbed, meaning that it does not
progress across the field. The second phase is an active one, where
the propagation of the virus reaches a macroscopic fraction of the
agricultural field. The transition from the absorbing to the active
phase strictly resembles a non-equilibrium second order phase transition
in statistical physics \cite{Transitions_book2}. The critical value
of this transition $\left(\frac{\beta}{\mu}\right)_{c}=\tau^{\,E}$
is defined as the \textit{epidemic threshold}. This term is also know
as the\textit{ basic reproduction number} and it represents a threshold
in the sense that below this point the propagation of the infection
dies out and over it the propagation becomes an epidemic. We have
then investigated the epidemic threshold $\tau^{\,E}$ for the monocrop
and the six intercrop arrangements without ($\gamma=1.0)$ and with
a trap crop ($\gamma=2.0)$. We also considered, as before, two kinds
of aphid, one with relatively low mobility ($s=4.0)$ and the other
with higher mobility ($s=2.5)$. In Fig. \ref{thresholds} we resume
the results. Let us first consider the intercropped fields without
trap crops ($\gamma=1.0$). Then, when the pest has low mobility the
epidemic threshold of the chessboard arrangement is more than 10 times
higher than that of the monocrop. Notice that we have normalized all
the bar plots in the insets of Fig. \ref{thresholds} by dividing
the epidemic thresholds by that of the monocrop. Indeed, we have proved
in the Supplementary Note 5 that the chessboard arrangement can reach
an infinitely large epidemic threshold if $s$ is bounded and the
trap crop has a very high strength. The rest of the arrangements have
epidemic thresholds which are about twice that of the monocrop. When
the aphid mobility increases, the epidemic thresholds logically drop,
due to the fact that it is easier for the pest to trigger the propagation
of a virus across the field. In this case the chessboard arrangement
triplicates the epidemic threshold of the monocrop, while the rest
of the arrangements have values of about 1.5 times larger than the
one of the monocrop. When we incorporate trap crops ($\gamma=2.0$)
in the intercrop arrangements the changes in the epidemic threshold
results very dramatic for the case of the chessboard arrangement.
In this case, with low and high mobility, the epidemic thresholds
are about 40 and 32 times higher than that of the monocrop. For the
rest of the intercropped systems the threshold increases by factors
between 2 and 5. It is interesting that for the rest of the intercrop
systems the ordering of the epidemic thresholds vary from one scenario
to another. For instance, without trap crop and low mobility of the
pest, the random arrangement is the second best, followed by the rows
arrangement. However, if the aphid has larger mobility the column
arrangement is the second best followed by the rows one. When there
are trap crops and low mobility of the pest the rows arrangement is
the second best followed by the columns one. If the mobility of the
pest is higher then the column is the second best followed by the
strips one. It is possible that the empirical observation that the
rows and strips arrangement delay the propagation of an aphid-borne
virus in a crop field has made that these two arrangements have been
the most widely used ones. However, in terms of (i) percentage of
plants removed by the infection, (ii) rate of the propagation of the
aphid-borne virus across the field, and (iii) epidemic threshold,
the chessboard arrangement introduced here is by far the most efficient
intercrop arrangement without and with trap crops. \textcolor{black}{In
this respect our model agrees with previous results showing that the
finer grained mixing of susceptible and resistant species impedes
the propagation of diseases on plants \cite{Mundt_1,Mundt_2,Skelsey 2005}.
However, as we have also shown in this work, when the radii of aphid
movement increases, then other intercropping arrangements such as
the column, rows and random, are very efficient to stop the propagation
of diseases across a field.}

\section*{Conclusion}

\textcolor{black}{Here, we demonstrate using intensive mathematical
modeling, that the efficiency of intercropping arrangements can be
improved dramatically in relation to the designs currently in use.
We develop a mathematical framework that allows to study the effect
of intercropping systems with and without 'trap crops'. Our study
shows that improving existing intercrop designs may decrease up to
80\% the number of plants affected by aphid-borne viruses, slow down
the propagation of such aphid-borne viruses by a 300-fold factor,
and delay the triggering of these epidemics on plants by a 40-fold
factor respect to a monocrop. Indeed, our analytical and numerical
findings show that the chessboard is the best arrangement when the
pest can hop only across the rows and columns, but not diagonally
between rows.}

\section*{Acknowledgements}

EE thanks inspirational encouragement from Consuelo Ramos-L\'opez who
has used intercropping for more than 60 years. AA-P acknowledges the
support of the Brazilian agency CNPq through Grant No. 151466/2018-1.

\section*{\textcolor{black}{\clearpage}}

\section*{\textcolor{black}{Supplementary Tables}}

\textcolor{black}{}
\begin{table}[H]
\begin{centering}
\textcolor{black}{}%
\begin{tabular}{|c|c|c|c|}
\hline 
\textcolor{black}{intercrop system} & \textcolor{black}{intercrop system} & \textcolor{black}{intercrop system} & \textcolor{black}{intercrop system}\tabularnewline
\hline 
\hline 
\textcolor{black}{alfalfa\_potato \cite{potato_alfalfa}} & \textcolor{black}{maize\_cowpea} & \textcolor{black}{maize\_mungbean} & \textcolor{black}{sorghum\_groundnut}\tabularnewline
\hline 
\textcolor{black}{barley\_pea} & \textcolor{black}{maize\_wheat} & \textcolor{black}{maize\_egusi melon} & \textcolor{black}{sorghum\_soybean}\tabularnewline
\hline 
\textcolor{black}{barley\_clover} & \textcolor{black}{maize\_faba beans} & \textcolor{black}{maize\_ alfalfa} & \textcolor{black}{sorghum\_cowpea}\tabularnewline
\hline 
\textcolor{black}{barley\_medic} & \textcolor{black}{maize\_soybean} & \textcolor{black}{mustard\_legume} & \textcolor{black}{soybean\_sunflower}\tabularnewline
\hline 
\textcolor{black}{barley\_faba} & \textcolor{black}{maize\_bean} & \textcolor{black}{oat\_faba beans} & \textcolor{black}{strawberry\_broad beans}\tabularnewline
\hline 
\textcolor{black}{barley\_vetch} & \textcolor{black}{maize\_bean\_squash} & \textcolor{black}{oat\_common vetch} & \textcolor{black}{sweet potato\_jugo beans}\tabularnewline
\hline 
\textcolor{black}{barley\_chickpea} & \textcolor{black}{maize\_canola} & \textcolor{black}{oat sown\_vetch} & \textcolor{black}{wheat\_faba beans}\tabularnewline
\hline 
\textcolor{black}{coffee\_macuaba \cite{coffee_palm}} & \textcolor{black}{maize\_tobacco} & \textcolor{black}{okra\_pumpkin} & \textcolor{black}{wheat\_soybean}\tabularnewline
\hline 
\textcolor{black}{cowpea\_soybean} & \textcolor{black}{maize\_sugarcane} & \textcolor{black}{okra\_sweet potato\_} & \textcolor{black}{wheat\_canola}\tabularnewline
\hline 
\textcolor{black}{cowpea\_cotton} & \textcolor{black}{maize\_potato} & \textcolor{black}{onion\_pepper} & \textcolor{black}{wheat\_chickpea}\tabularnewline
\hline 
\textcolor{black}{faba beans\_pea} & \textcolor{black}{maize\_peanut} & \textcolor{black}{radish\_amarantus} & \textcolor{black}{wheat\_common vetch}\tabularnewline
\hline 
\textcolor{black}{faba\_teff} & \textcolor{black}{maize\_legume} & \textcolor{black}{rice\_blackgram} & \textcolor{black}{wheat\_lentils}\tabularnewline
\hline 
\textcolor{black}{fennel\_dill} & \textcolor{black}{maize\_okra} & \textcolor{black}{sesame\_sunflower} & \textcolor{black}{wheat\_cotton}\tabularnewline
\hline 
\textcolor{black}{leek\_celery} & \textcolor{black}{maize\_groundnut} & \textcolor{black}{sorghum\_legume} & \tabularnewline
\hline 
\textcolor{black}{lentils\_fenu greek} & \textcolor{black}{maize\_peanut} & \textcolor{black}{sorghum\_sunflower} & \tabularnewline
\hline 
\end{tabular}
\par\end{centering}
\textcolor{black}{\caption{Illustration of some intercropped systems reported by Brookers et
al. \cite{Brooker-1}, Martin-Guay et al. \cite{Martin-Guay-1} and
references therein.}
}
\end{table}

\section*{\textcolor{black}{Supplementary Note 1}}

\textcolor{black}{In formulating the ``rationale of the model''
we consider that two plants $i$ and $j$ are separated by a geographic
(Euclidean) distance equal to $\rho_{ij}$. We consider that the aphid
has a radius of primary exploration equal to $r$. That is, the aphid
can hop directly from the plant $i$ to any other plant in a radius
$r$ from $i$. Then, if two plants are at a distance $\rho_{ij}\leq r$
we consider that they are topologically connected to each other by
an edge. The number of edges in the shortest path connecting two plants
$i$ and $k$ (not directly connected to each other) is the topological
distance $d_{ik}.$ We consider here two different kinds of hopping
probabilities. The geographic hopping probability $\pi_{ij}$ depends
on the geographic distance between the two plants, e.g., $\pi_{ij}\propto\rho_{ij}^{-s}$.
The topological hopping probability depends only on the topological
distance separating the two plants, and not on its geographic separation,
e.g., $p_{ij}\propto d_{ij}^{-s}$. Then, we have the following rules.}

\textbf{\textcolor{black}{First come first served}}\textcolor{black}{.
Consider an aphid at a plant $i$ which can hop to any of the adjacent
plants $j$ (left) or $k$ (right) (see Supplementary Figure \ref{First come}(a)).
Let us consider that the geographic distance between the plants are
$\rho_{ij}<r$ and $\rho_{ik}<r$, respectively, such that $\rho_{ij}>\rho_{ik}$.
Then, according to the geographic distance, the probability of the
aphid hopping to plant $k$ is larger than that of hopping to plant
$j$, i.e., $\pi_{ij}\propto\rho_{ij}^{-s}$ and $\pi_{ik}\propto\rho_{ik}^{-s}$
assuming a power-law decay with the distance, thus $\pi_{ij}<\pi_{ik}$.
However, as the plants $j$ and $k$ are inside the radius of primary
movement of the aphid, an aphid hopping from $i$ to the right will
find first the plant $k$, exactly the same as an aphid hopping to
the left who will find first the plant $j$. Thus, both hopping processes
should display the same probabilities, which is accounted for by the
topological distance between the pairs of plants. Because $\left(i,j\right)$
are nearest neighbors as well as $\left(i,k\right)$, we have that
$d_{ij}=d_{ik}=1$, where $d_{ij}$ is the topological (shortest path)
distance. Consequently, $p_{ij}=p_{ik}$. operators}

\begin{figure}[H]
\begin{centering}
\textcolor{black}{\includegraphics[width=1\textwidth]{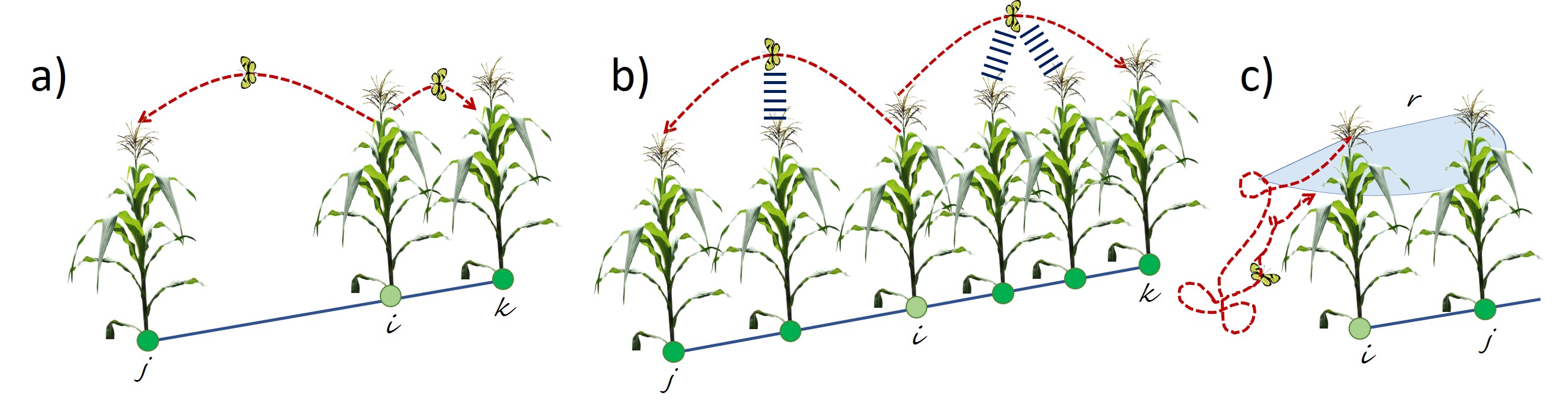}}
\par\end{centering}
\caption{(a) Illustration of the ``first come first served'' hopping process
of an aphid between plants. (b) Illustration of the exploratory strategy
of aphid based on ``a bird in the hands is worth than two in the
bush''. (c) Illustration of the plausible returning strategy of an
aphid after overpassing its exploratory radius without finding other
plants to land in.}

\label{First come}
\end{figure}

\textbf{\textcolor{black}{A bird in the hand is worth more than two
in the bush}}\textcolor{black}{. Consider an aphid at a plant $i$
which can hop either to plants on its left or on its right (see Supplementary
Figure \ref{First come} (b)). Let us consider that the geographic
distances between the plants are $\rho_{ij}$=$\rho_{ik}>r$. Then,
according to the geographic distance, the probability of the aphid
to hop to plant $k$ is exactly the same as that of hopping to plant
$j$, i.e., $\pi_{ij}=\pi_{ik}\propto\rho_{ij}^{-s}$ assuming a power-law
decay with the distance. However, an aphid moving from $i$ to the
left will find first a plant in its way to $j$. Thus, assuming that
such plant is attractive to it, the aphid will explore first that
plant on the basis of a minimum effort principle and then the plant
$j$. On the other hand, an aphid moving from $i$ to the right will
find first a plant that it can explore and in case it decides to continue
its movement to the right, the aphid will find yet another plant before
arriving at $k$. Thus, it is clear that under the same conditions
the probability of arriving at the plant $k$ is smaller than that
of arriving at $j$, although they are at exactly the same geographic
distances. Assuming the connectivity of the plants given in Supplementary
Figure \ref{First come} (b) we have that $2^{-s}\propto p_{ij}>p_{ik}\propto3^{-s}$.}

\textbf{\textcolor{black}{Go back before it is too late}}\textcolor{black}{.
Consider an aphid having an exploratory movement at the borderline
of a crop field from a node $i$ (see Supplementary Figure \ref{First come}(c)).
If the aphid is moving away from the field there is a high probability
that it overpass its exploratory radius before finding a new plant.
Thus, it is probable that the aphid returns to plant $i$ before finding
any new one. As a consequence, the probability that the aphid arrives
at a plant distant from $i$ depends more on the topological separation
among the plants than on the geographic distance through the ``possibly
empty'' space separating them. That is, we consider here that an
aphid navigates a crop field by orienting itself through the plants
and not realizing ``risky'' explorations outside the field.}

\section*{\textcolor{black}{Supplementary Note 2}}

\noindent \textcolor{black}{More formally, $\tilde{A}$ is a transformed
adjacency operator on a graph, which will be defined as follows. Let
us consider $\Gamma=(V,E)$ to be an undirected finite or infinite
graph with vertices $V$ and edges $E$. We assume that $\Gamma$
is connected and locally finite (i.e.\ each vertex has only finitely
many edges emanating from it). Let $d$ be the shortest path distance
metric on $\Gamma$, i.e.\ $d(v,w)$ is the length of the shortest
path from $v$ to $w$. Let $\ell^{2}(V)$ be the Hilbert space of
square-summable functions on $V$ with inner product 
\begin{align}
\langle f,g\rangle=\sum_{v\in V}f(v)\overline{g(v)},\qquad f,g\in\ell^{2}(V).
\end{align}
In $\ell^{2}(V)$ there is a standard orthonormal basis consisting
of the vectors $e_{v}$, $v\in V$, where 
\begin{equation}
e_{v}(w)=\begin{cases}
1 & \text{if \ensuremath{w=v}},\\[0.5ex]
0 & \text{otherwise}.
\end{cases}\label{defev}
\end{equation}
}

\textcolor{black}{For $d\in\mathbb{N}$ the following operator defined
in $\ell^{2}(V)$ is the }\textit{\textcolor{black}{$d$-path adjacency
operator of the graph}}\textcolor{black}{{} 
\begin{equation}
\bigl(A_{d}f\bigr)(v)\coloneqq\sum_{w\in V:\,d(v,w)=d}\bigl(f(w)\bigr),\qquad f\in\ell^{2}(V),v\in V\label{eq:path_adjacency}
\end{equation}
}

\textcolor{black}{The $d$-path adjacency operator acts over the vectors
$e_{v}$ as}

\textcolor{black}{
\begin{equation}
(A_{d}e_{v})(w)=\begin{cases}
1 & \text{if \ensuremath{d(v,w)=d}},\\[0.5ex]
0 & \text{otherwise}.
\end{cases}\label{Lkev}
\end{equation}
}

\textcolor{black}{These operators are the adjacency analogues of the
}\textit{\textcolor{black}{$d$-path Laplacian operators }}\textcolor{black}{of
the graph \cite{path Laplacian 1,path Laplacian 2,path Laplacian 3}.
The Mellin (power-law) transformed adjacency operator is then defined
by}

\textcolor{black}{
\begin{equation}
\tilde{A}\coloneqq\sum_{d=1}^{d_{max}}d^{-s}A_{d}.\label{deftLM}
\end{equation}
}

\textcolor{black}{Other transforms are also possible as the Laplace
(exponential) one (see for instance \cite{path Laplacian 1,path Laplacian 2,path Laplacian 3}
for the analogues in the path Laplacians), but we constraint ourselves
here to the power-law one.}

\textcolor{black}{For instance, if we consider a rectangular lattice
like the one illustrated in the Supplementary Figure \ref{rectangular}
in which we have labeled the nodes using numbers from 1 to 12. The
corresponding Mellin transformed $d$-path adjacency matrix for this
network is given below for a generic value of the insect mobility
$s$:}
\begin{center}
\textcolor{black}{$\tilde{A}=\left(\begin{array}{cccccccccccc}
0 & 1 & 2^{-s} & 3^{-s} & 1 & 2^{-s} & 3^{-s} & 4^{-s} & 2^{-s} & 3^{-s} & 4^{-s} & 5^{-s}\\
 & 0 & 1 & 2^{-s} & 2^{-s} & 1 & 2^{-s} & 3^{-s} & 3^{-s} & 2^{-s} & 3^{-s} & 4^{-s}\\
 &  & 0 & 1 & 3^{-s} & 2^{-s} & 1 & 2^{-s} & 4^{-s} & 3^{-s} & 2^{-s} & 3^{-s}\\
 &  &  & 0 & 4^{-s} & 3^{-s} & 2^{-s} & 1 & 5^{-s} & 4^{-s} & 3^{-s} & 2^{-s}\\
 &  &  &  & 0 & 1 & 2^{-s} & 3^{-s} & 1 & 2^{-s} & 3^{-s} & 4^{-s}\\
 &  &  &  &  & 0 & 1 & 2^{-s} & 2^{-s} & 1 & 2^{-s} & 3^{-s}\\
 &  &  &  &  &  & 0 & 1 & 3^{-s} & 2^{-s} & 1 & 2^{-s}\\
 &  &  &  &  &  &  & 0 & 4^{-s} & 3^{-s} & 2^{-s} & 1\\
 &  &  &  &  &  &  &  & 0 & 1 & 2^{-s} & 3^{-s}\\
 &  &  &  &  &  &  &  &  & 0 & 1 & 2^{-s}\\
 &  &  &  &  &  &  &  &  &  & 0 & 1\\
 &  &  &  &  &  &  &  &  &  &  & 0
\end{array}\right)$}
\par\end{center}

\begin{figure}[H]
\begin{centering}
\textcolor{black}{\includegraphics[width=0.45\textwidth]{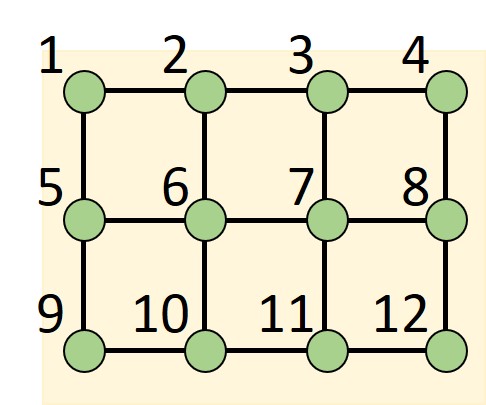}}
\par\end{centering}
\caption{Illustration of a rectangular lattice with nodes labeled.}

\label{rectangular}
\end{figure}

\section*{\textcolor{black}{Supplementary Note 3}}

\textcolor{black}{We consider the situation in which the insect pest
has a radius for short-range dispersal\textendash searching, foraging,
ranging, etc.\textendash which is larger than the one considered in
the main text. The influence of the inter-rows separation was previously
considered by Sheoran et al. experimentally \cite{rows_2-1}. However,
their analysis is only from the bio-economical point of view and not
from that of pest management. Here we consider $r=\sqrt{2}\varDelta$
instead of $r=\varDelta$, which gives rise to networks like the one
illustrated in the Supplementary Figure \ref{longer radius}. This
situation naturally emerges if we consider that the rows and columns
of the crop field are too close together, such that the pest can hop
not only across the rows and columns, but also diagonally between
rows. To notice the difference between this network and the square
one produced by $r=\varDelta$ we simply should count the number of
steps a pest need to traverse the plot through its diagonal. In the
square lattice the number of steps needed is $D=c+r-2$, where $c$
is the number of columns, and $r\leq c$ is the number of rows. However,
in the lattice emerging from $r=\sqrt{2}\varDelta$ we have $D=c-1$,
which is always smaller that the previous one if $c>1$. This difference
is important as it will provide strategies to mitigate the pest damages
produced in this kind of arrangements.}

\begin{figure}[H]
\begin{centering}
\textcolor{black}{\includegraphics[width=0.45\textwidth]{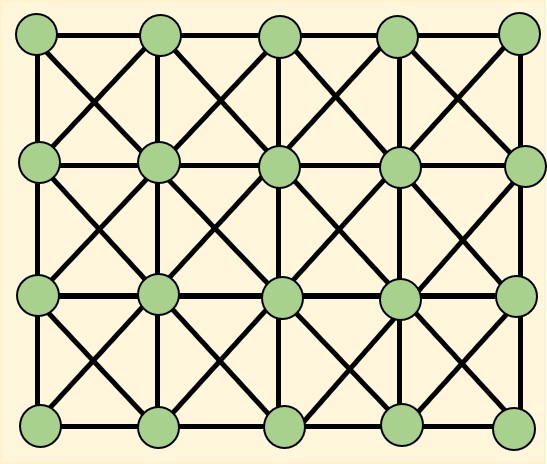}}
\par\end{centering}
\caption{Network obtained for a monocrop in which the radius of foraging of
the insect pest is $r=\sqrt{2}\varDelta$ instead of $r=\varDelta$.}

\label{longer radius}
\end{figure}

\textcolor{black}{In the Supplementary Figure \ref{longer radius no trap}
we give the results for the simulations of a pest propagation in these
crop fields when there are no trap crops ($\gamma=1.0$). When the
pest mobility is relatively low ($s=4.0$) the best arrangements are
the rows and columns intercrops with about 10\% of affected plants
vs. 78\% affected in the monocrop. In this case the chessboard and
random intercrops have 30\% of affected plants and are the second
best, followed by strip (31\%), and patches (57\%). The rate of propagation
$v$ indicates that pests propagating across a monocrop infestates
29.4 plants per unit of time, vs. 9 plants infected in the rows and
column arrangements, and 10 plants in the random or 12 in the chessboard.
The situation is considerably worsened when the pest has high mobility
($s=2.5$). In this case none of the intercropping systems is able
to stop the propagation of the pest across the field with percentages
of affected plants ranging 93-98\%, not different from that in the
monocrop (98.6\%). The rates of propagation are reduced to about a
half relative to that in the monocrop (37 plants infected per unit
time). The reason for this catastrophic result is that a pest with
relatively high mobility can infestate very quickly large neighborhoods
of a plant due to the possibility that it has of infecting plants
in any direction from its current position. An obvious measure to
mitigate this problem is to increase the separation of the rows and
columns in the crop field, or even\textendash as shown in the experiments
by Khan et al. \cite{Kahn_et_al}\textendash to increase the separation
between rows keeping a smaller separation between columns.}

\begin{figure}[H]
\textcolor{black}{}\subfloat[]{\textcolor{black}{\includegraphics[width=0.45\textwidth]{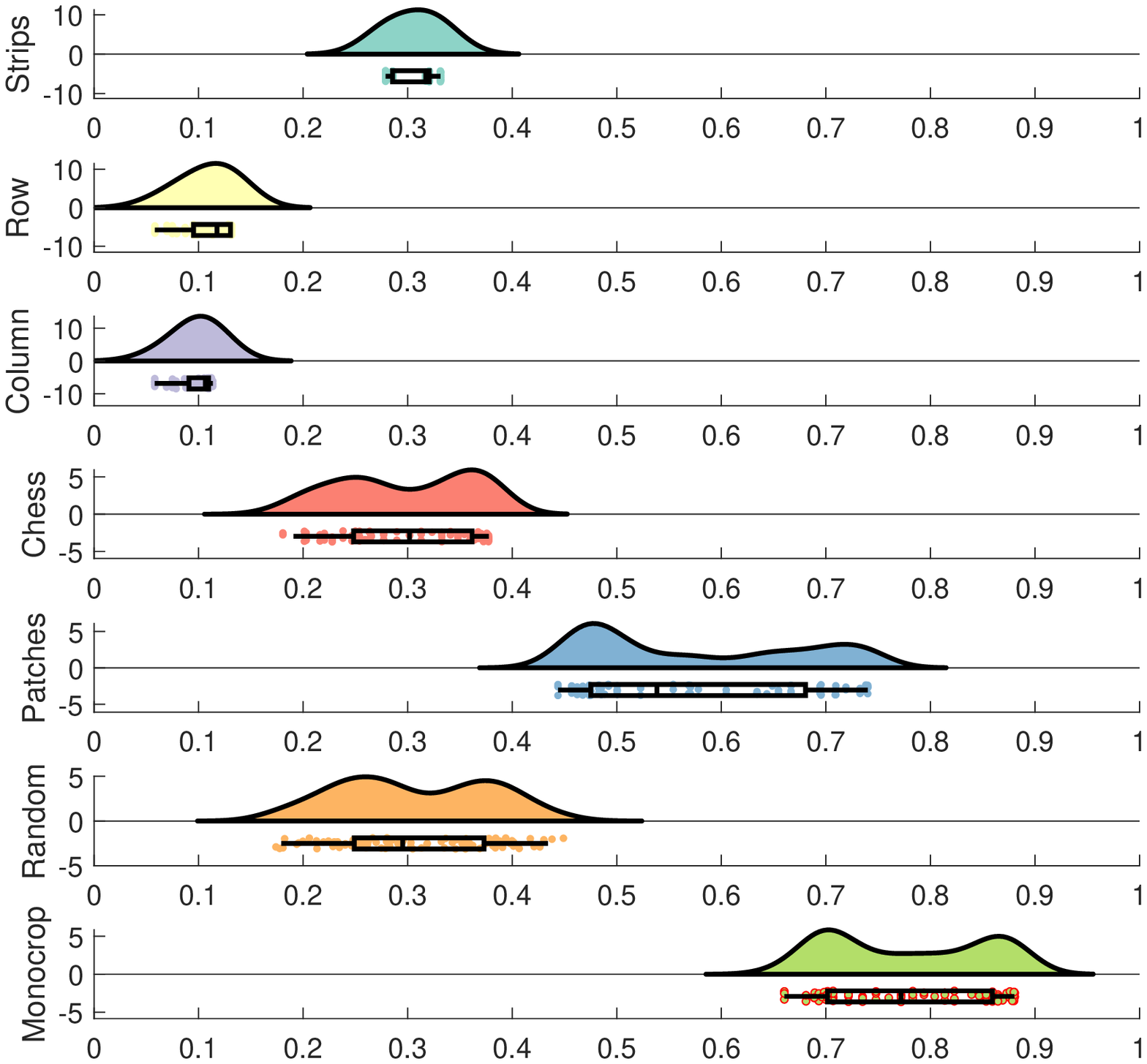}}

\textcolor{black}{}}\textcolor{black}{}\subfloat[]{\textcolor{black}{\includegraphics[width=0.45\textwidth]{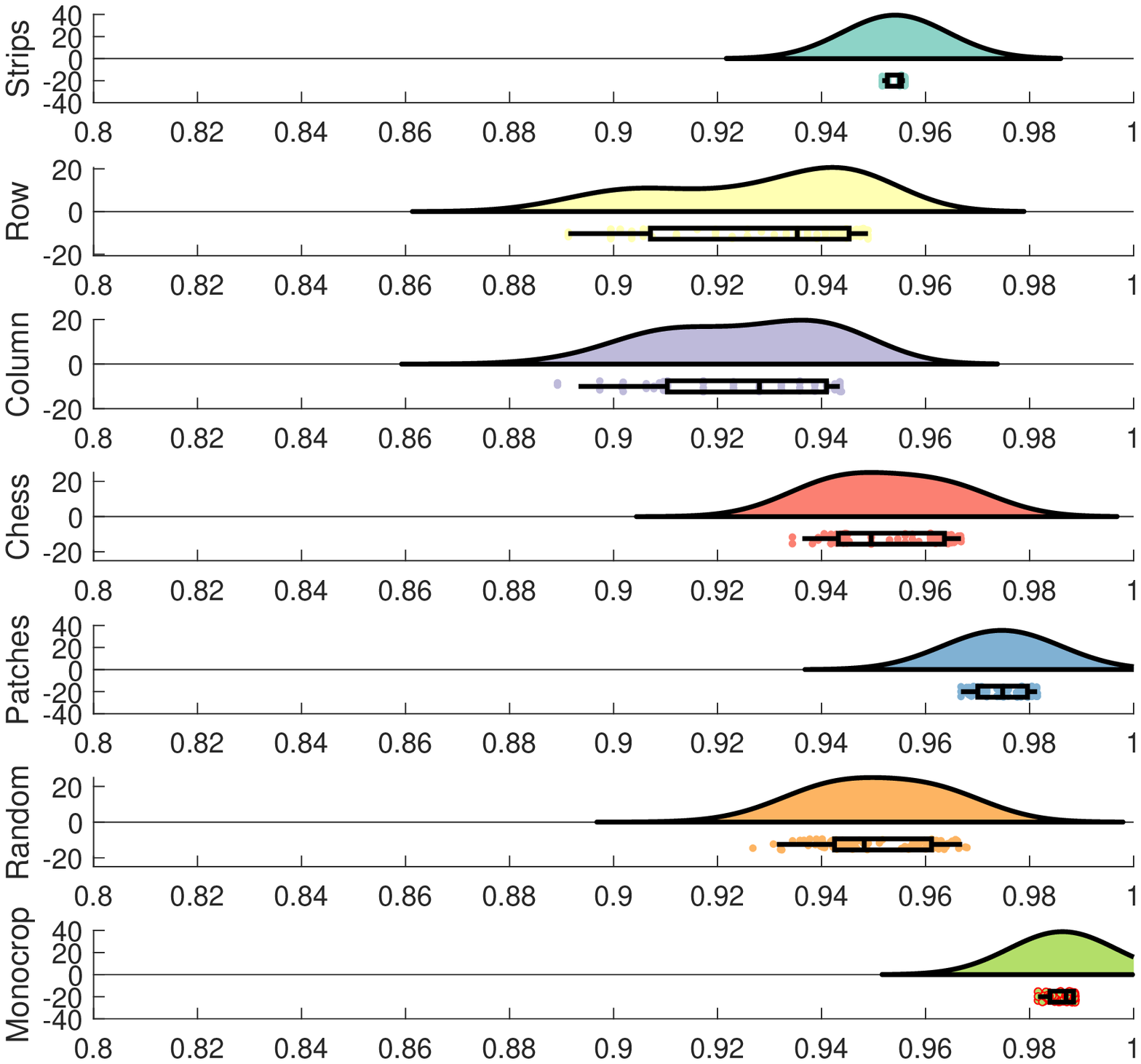}}

\textcolor{black}{}}

\textcolor{black}{}\subfloat[]{\textcolor{black}{\includegraphics[width=0.45\textwidth]{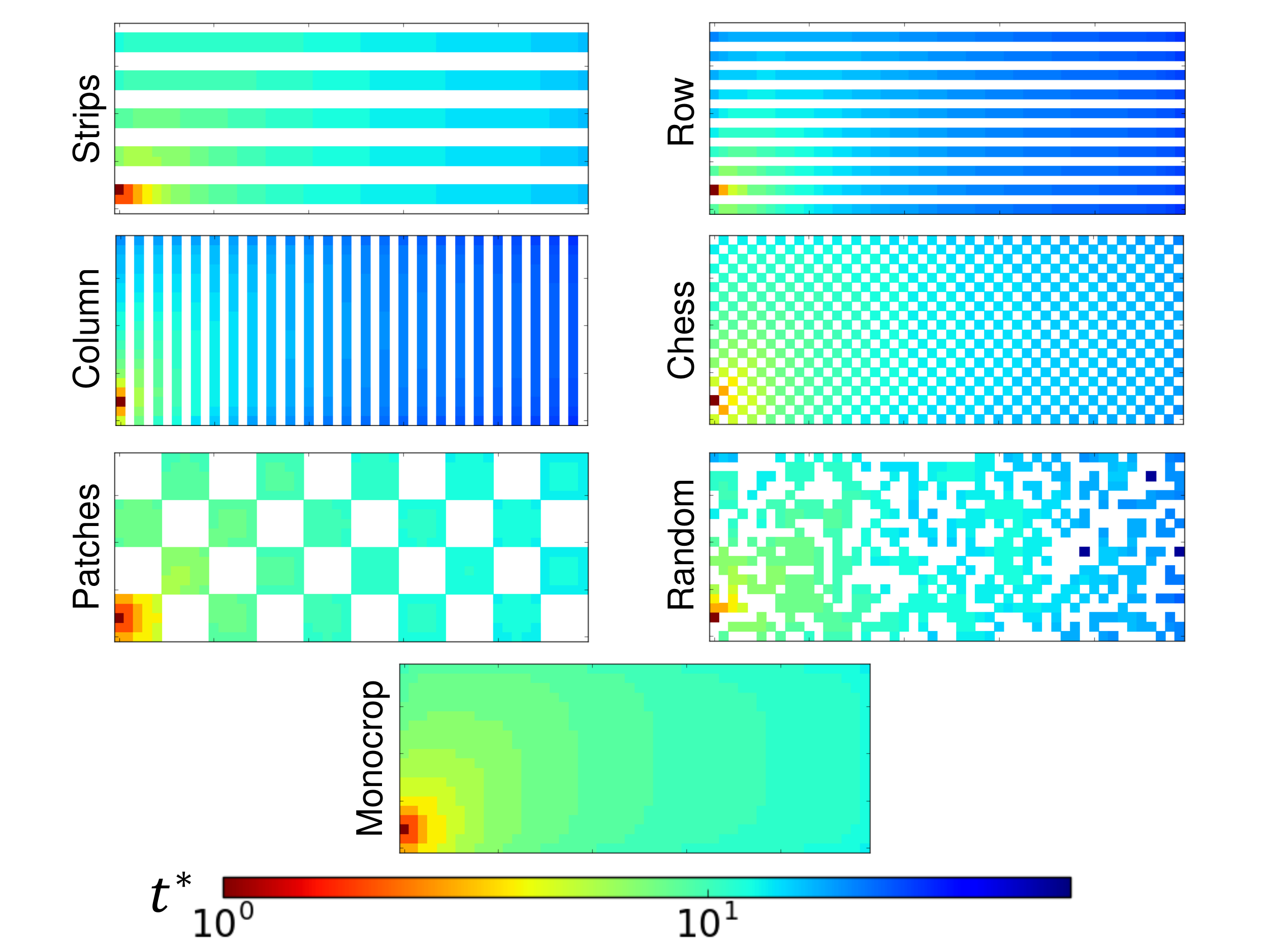}}

\textcolor{black}{}}\textcolor{black}{}\subfloat[]{\textcolor{black}{\includegraphics[width=0.45\textwidth]{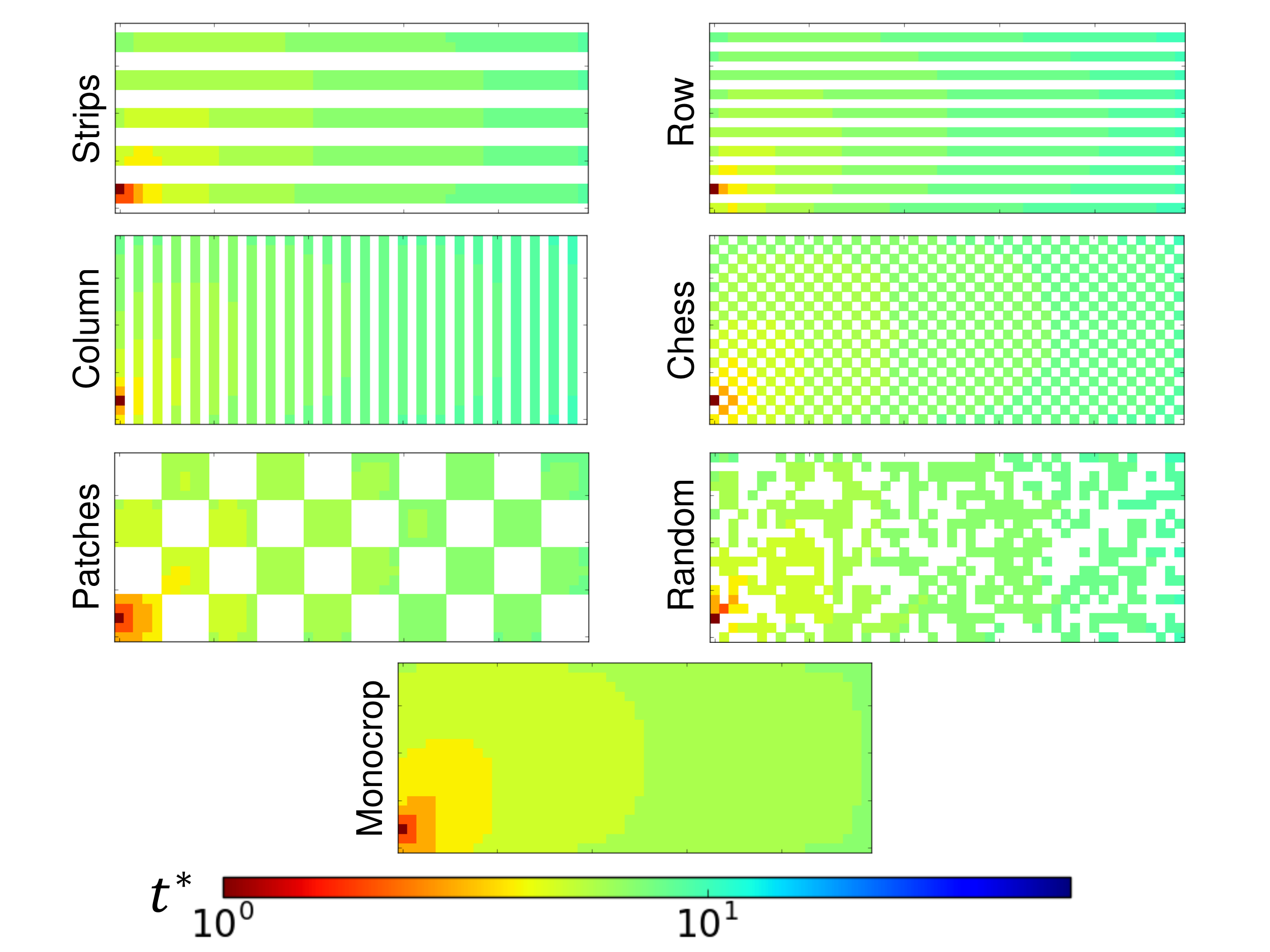}}

\textcolor{black}{}}

\caption{Results of the simulations for a SIR epidemics at $t=10$ with $r=\sqrt{2}\varDelta$,
$\beta=0.5,$ $\mu=0.5$ for different intercropping strategies without
traps ($\gamma=1.0$). (a) Raincloud plot of the proportion of dead
plants for pest with reduced mobility ($s=4.0)$. (b) Raincloud plot
of the proportion of dead plants for pest with relatively high mobility
($s=2.5$). The clouds show the kernel distribution of the proportion
of dead plants for different realizations of the epidemics. Below,
the raw data is plotted (the rain) together with their corresponding
box and whisker plots. (c) Evolution of the propagation of a relatively
low mobility pest ($s=4.0)$ across the fields. (d) Evolution of the
propagation of a relatively high mobility pest ($s=2.5)$ across the
fields. In both cases the propagation is initialized by infecting
the plant on the bottom-right corner of the plot.}

\label{longer radius no trap}
\end{figure}

\textcolor{black}{When the trap crops ($\gamma=2.0$) are implemented
in the different arrangements the situation resembles more the results
reported in the main paper (see Supplementary Figure \ref{longer radius trap}).
When the pest has low mobility ($s=4.0)$ the percentage of affected
plants in the monocrop is 77\% while those in the columns and rows
intercrops are 1.4\% and 1.9\%, respectively, followed very closely
by the chessboard arrangement with 2.1\% of affected plants. When
the trap crops are implemented the percentage of plants affected reaches
its minimum for the random arrangement (2.1\%) followed by columns
(4.0\%), rows (5.8\%), patches (6.2\%) and chessboard (6.4\%). Only
the strips intercrop has higher percentage of affected plants (26.2\%).
The rates of propagation of the pest follows similar patterns as the
one observed for the percentage of affected plants.}

\begin{figure}[H]
\textcolor{black}{}\subfloat[]{\textcolor{black}{\includegraphics[width=0.45\textwidth]{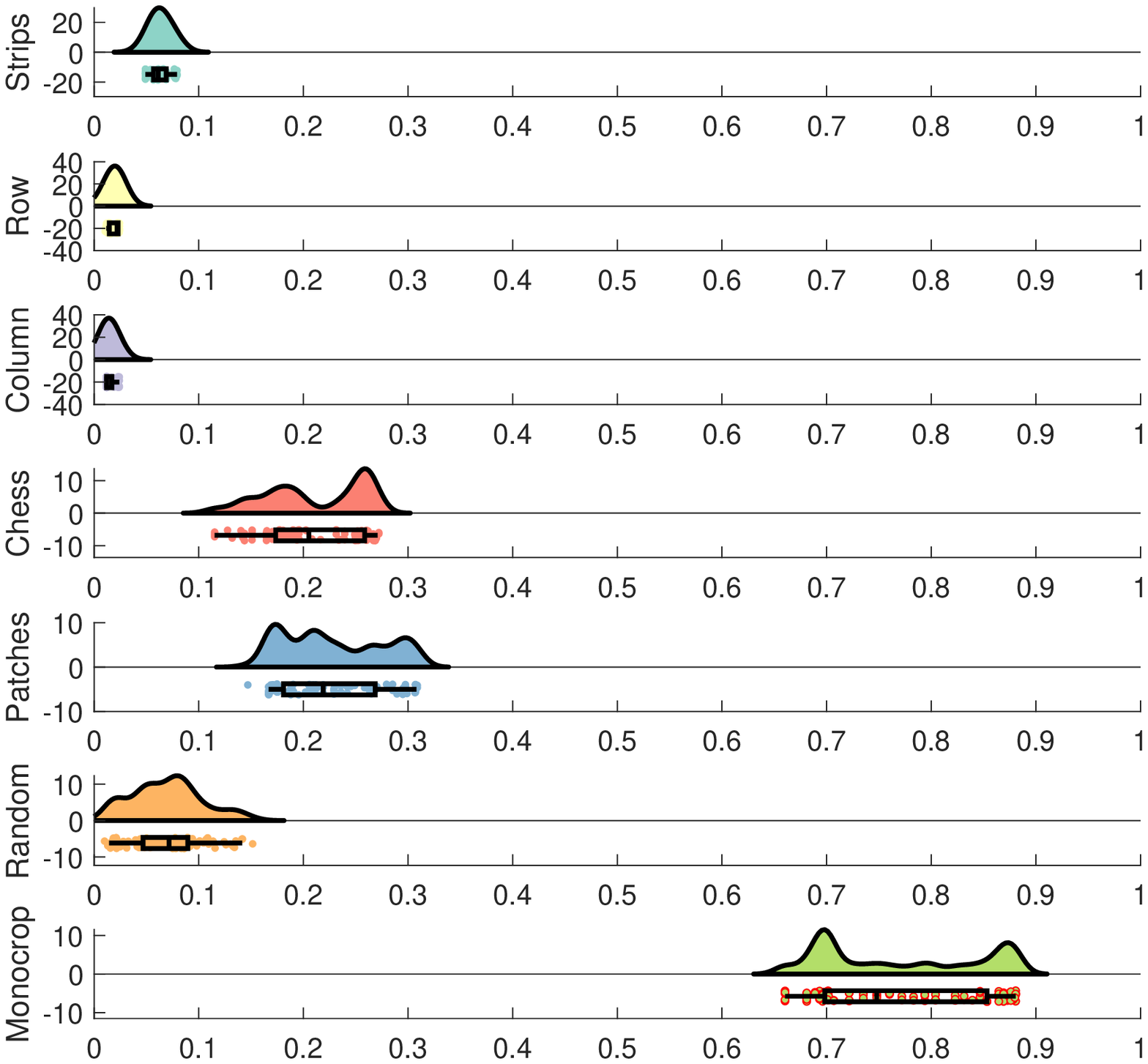}}

\textcolor{black}{}}\textcolor{black}{}\subfloat[]{\textcolor{black}{\includegraphics[width=0.45\textwidth]{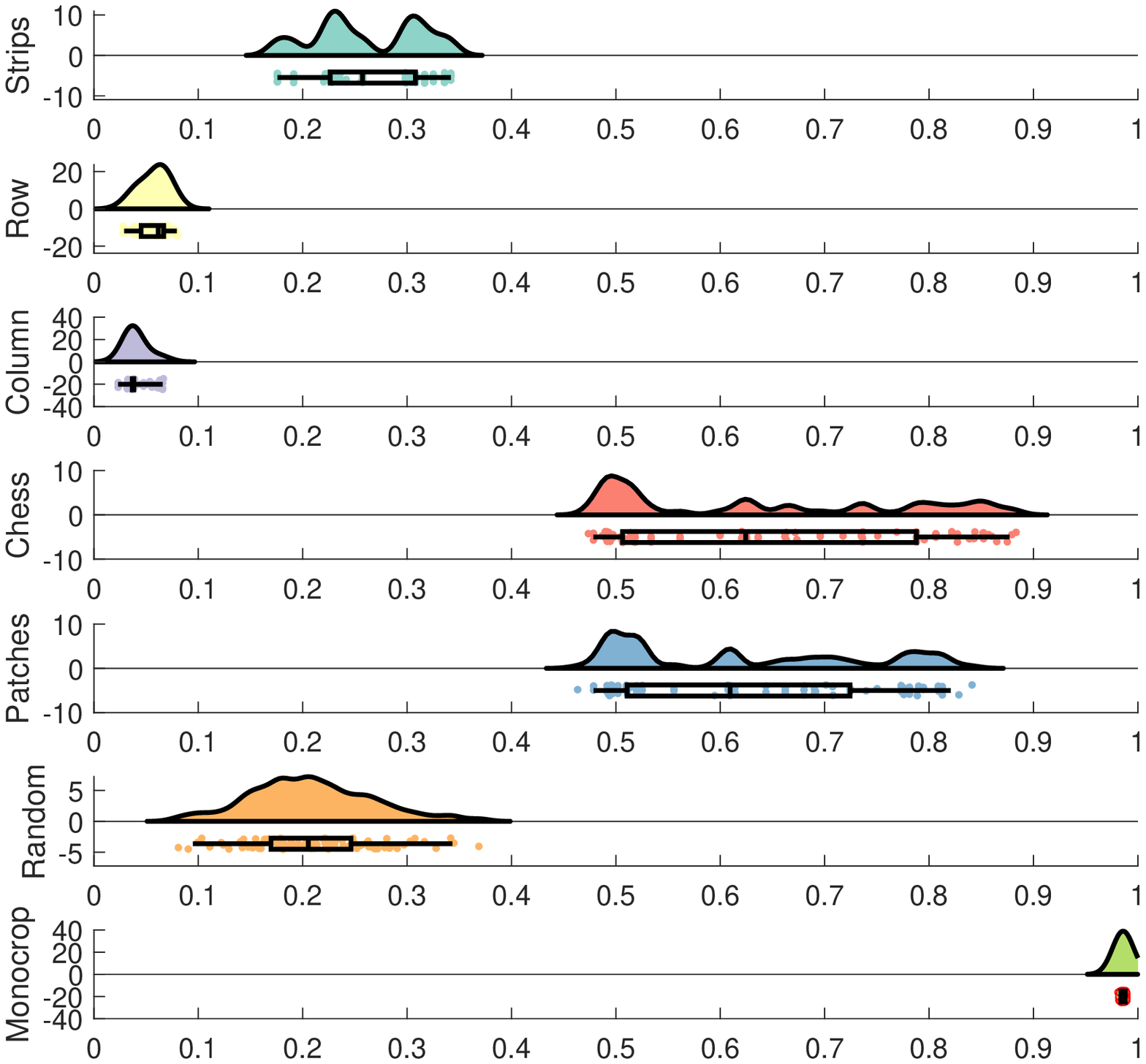}}

\textcolor{black}{}}

\textcolor{black}{}\subfloat[]{\textcolor{black}{\includegraphics[width=0.45\textwidth]{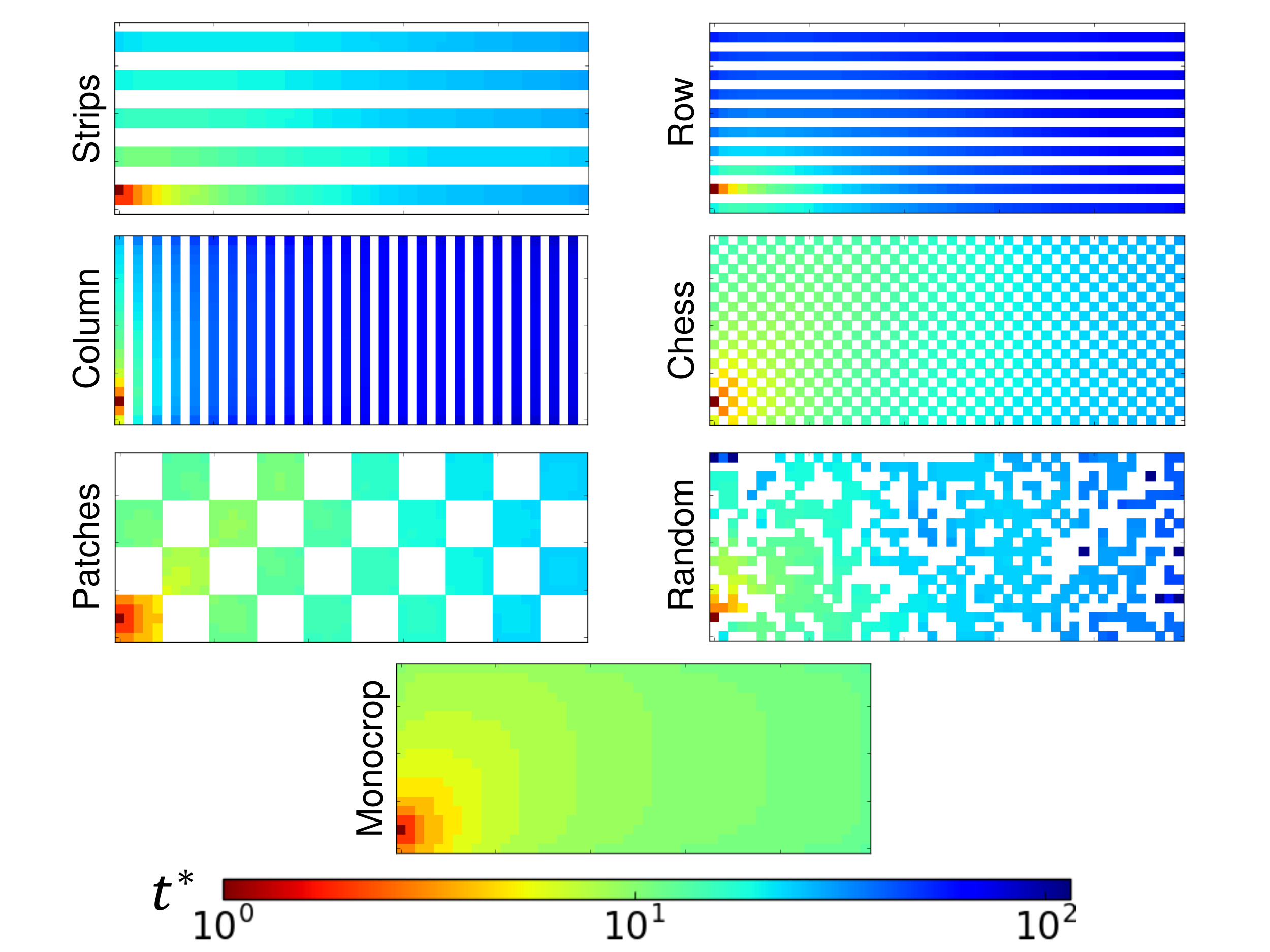}}

\textcolor{black}{}}\textcolor{black}{}\subfloat[]{\textcolor{black}{\includegraphics[width=0.45\textwidth]{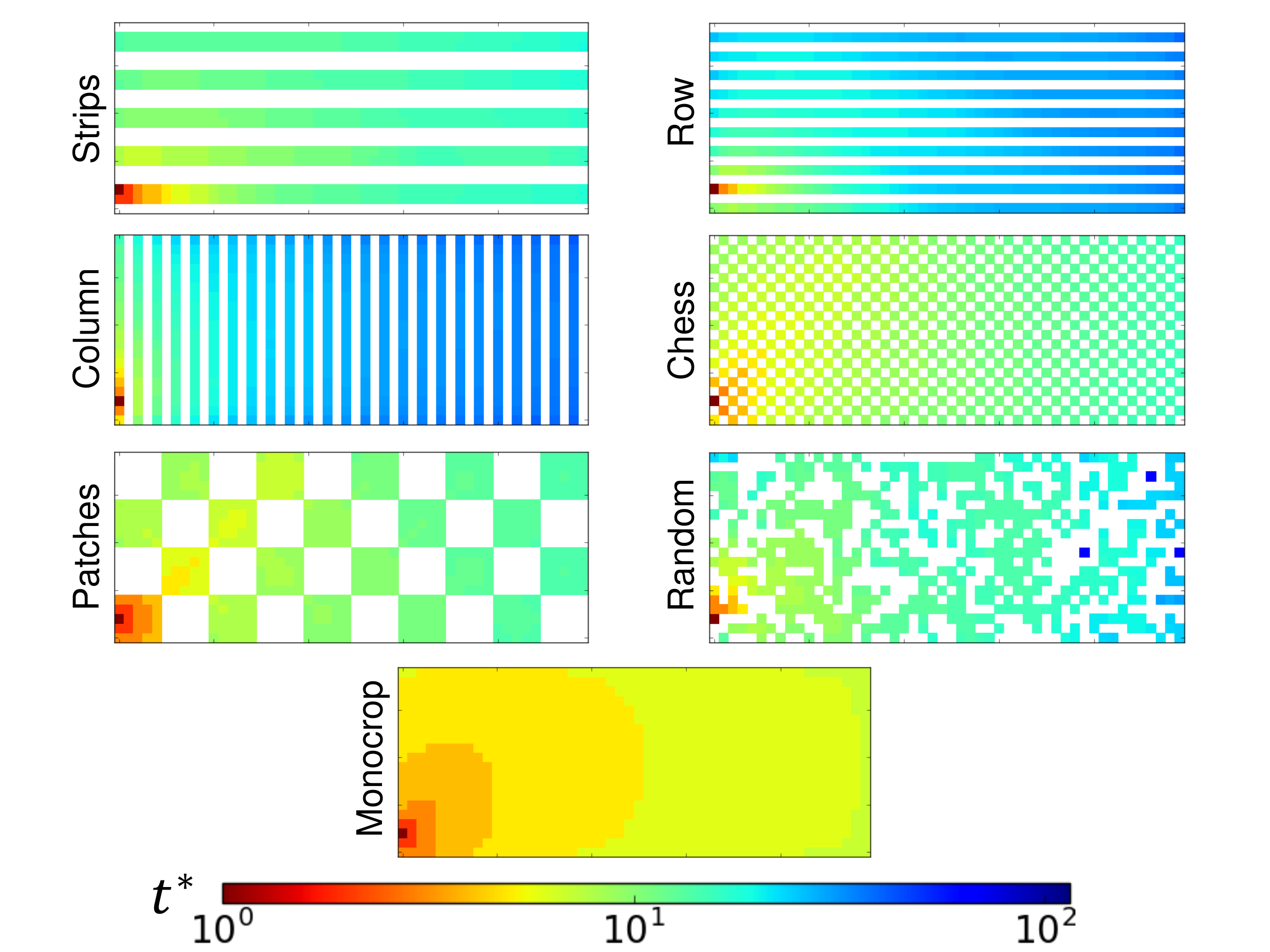}}

\textcolor{black}{}}

\caption{Results of the simulations for a SIR epidemics at $t=10$ with $r=\sqrt{2}\varDelta$,
$\beta=0.5,$ $\mu=0.5$ for different intercropping strategies with
trap crop of strength $\gamma=2$. (a) Raincloud plot of the proportion
of dead plants for pest with reduced mobility ($s=4.0)$. (b) Raincloud
plot of the proportion of dead plants for pest with relatively high
mobility ($s=2.5$). The clouds show the kernel distribution of the
proportion of dead plants for different realizations of the epidemics.
Below, the raw data is plotted (the rain) together with their corresponding
box and whisker plots. (c) Evolution of the propagation of a relatively
low mobility pest ($s=4.0)$ across the fields. (d) Evolution of the
propagation of a relatively high mobility pest ($s=2.5)$ across the
fields. In both cases the propagation is initialized by infecting
the plant on the bottom-right corner of the plot.}

\label{longer radius trap}
\end{figure}

\textcolor{black}{The most important conclusion of this Supplementary
Note is that the separation between rows and columns in the crop fields
should guarantee that the ``trivial'' dispersal of the pest should
be reduced as much as possible. Otherwise, the use of intercropping
without trap crops is very inefficient when the pest mobility is relatively
high. In any case, the use of trap crops continues to be the most
effective approach even in the case when the rows/columns separation
is not large enough as to limit the pest ``trivial'' dispersal.}

\section*{\textcolor{black}{Supplementary Note 4}}

\textcolor{black}{We consider intercrop systems with ``perfect''
trap crops. That is, we use $\gamma\rightarrow\infty$ for the two
cases previously analyzed of relatively low ($s=4.0$) and relatively
large ($s=2.5$) pest mobility. The result of this scheme is that
the pest cannot hop from one susceptible plant to another if in the
shortest path connecting them there is at least one trap. In the Supplementary
Figure \ref{Traps-1} we illustrate the results of our simulations
for these systems using the different arrangements studied here. As
can be seen for the case of relatively low mobility ($s=4.0$) there
are significant reduction in the percentages of affected plants for
all intercrop systems. The percentages of affected plants for each
intercrop are: chessboard (0.2\%), columns (1.3\%), random (1.5\%),
rows (1.7\%), strips (3.4\%) and patches (4.6\%). We remind the reader
that the percentage of affected plants in the monocrop is 18.1\%.
When the pest has a relatively large mobility ($s=2.5$) 95.1\% of
plants are affected in the monocrop, while in each of the intercrops
they are: chessboard (0.2\%), random (1.8\%), columns (2.5\%), rows
(3.9\%), patches (4.8\%), and strips (13.6\%). Notice that here there
are some important changes in the order of the arrangements in terms
of their effectivity in reducing the propagation of the pest. When
the pest is of high mobility the best arrangements are the chessboard
and the random one. The worse arrangement, and the only one having
more than 10\% of affected plants, is the strip one. Also notice that
the percentage of affected plants in the chessboard arrangement is
exactly the same for $s=2.5$ and $s=4.0$, indicating a high stability
in the efficiency of this arrangement. It is important to remark once
more time that these reductions in the number of affected plants are
the consequence of the different topological patterns emerging from
the different intercrop arrangements and not a dilution effect, as
the number of susceptible and immune plants are kept the same in every
arrangement.}

\textcolor{black}{We now analyze the rate of propagation of the insect
pests across the agricultural fields intercropped with a trap crop
(see Supplementary Figure \ref{Traps-1} (c) and (d)). Here the rate
of propagation of the insect pests follow a different order as for
the case of intercrops without trap crops. That is, for $s=4.0$,
we have: chessboard (0.05) < columns (0.34) < rows (0.70) < patches
(0.81) < random (1.41) < strips (2.04). For $s=2.5$, chessboard (0.05)
< columns (0.46) < patches (0.92) < rows (1.04) < random (1.66) <
strips (2.78). That is, here once more the chessboard arrangement
significantly outperforms the rest of intercrop arrangements. It is
also noticeable that the column arrangement is better than the rows
one in both cases, i.e., $s=2.5$ and $s=4.0$, and that the rows
intercrop can be even worse than the patches one if the pest has high
mobility. In this scenario the strips arrangement\textendash which
is frequently used in real-life intercrops\textendash is significantly
worse than the rest of the arrangements.}

\begin{figure}[H]
\begin{centering}
\textcolor{black}{}\subfloat[]{\begin{centering}
\textcolor{black}{\includegraphics[width=0.47\textwidth]{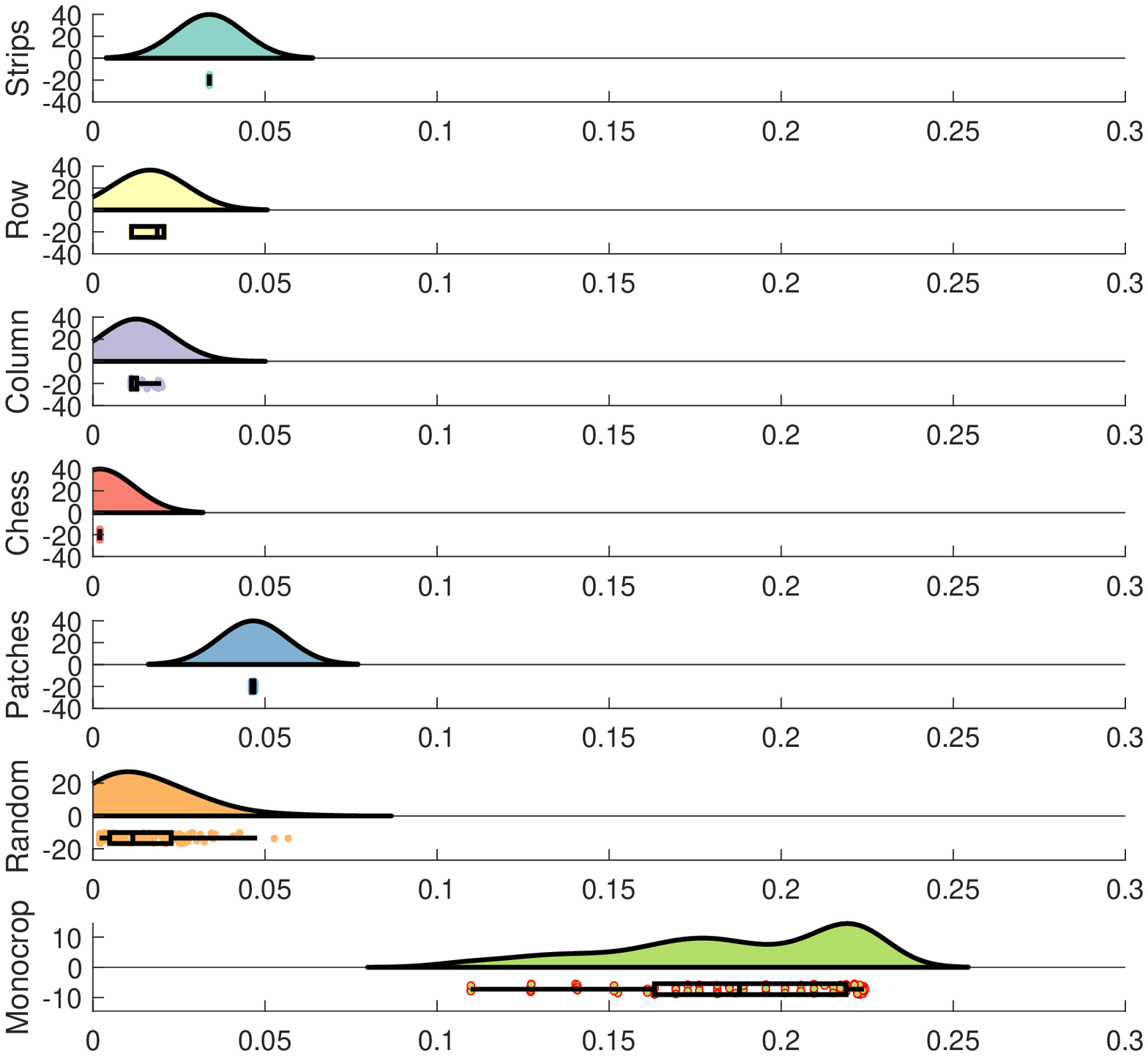}}
\par\end{centering}
\textcolor{black}{}}\textcolor{black}{}\subfloat[]{\begin{centering}
\textcolor{black}{\includegraphics[width=0.47\textwidth]{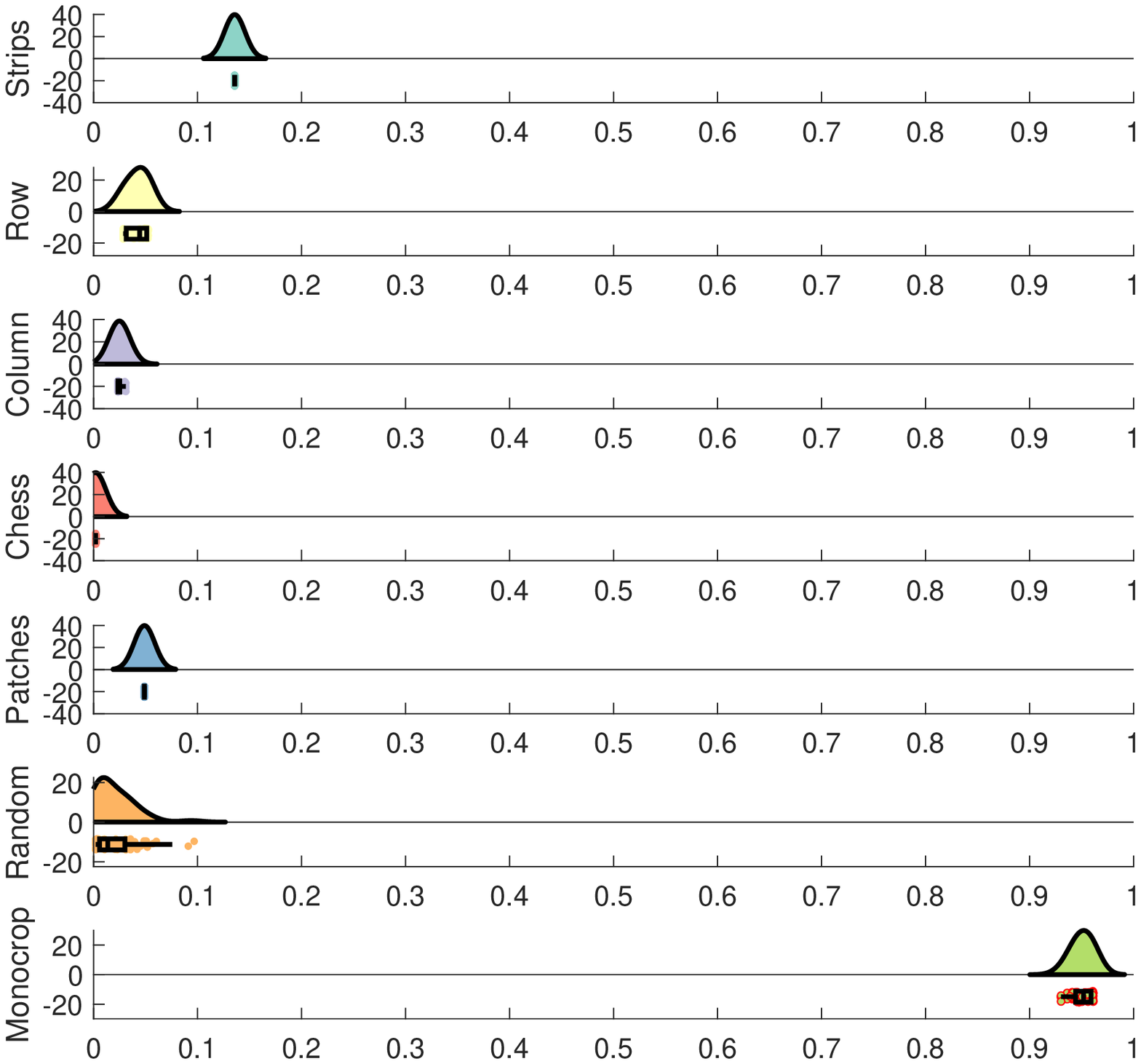}}
\par\end{centering}
\textcolor{black}{}}
\par\end{centering}
\begin{centering}
\textcolor{black}{}\subfloat[]{\begin{centering}
\textcolor{black}{\includegraphics[width=0.47\textwidth]{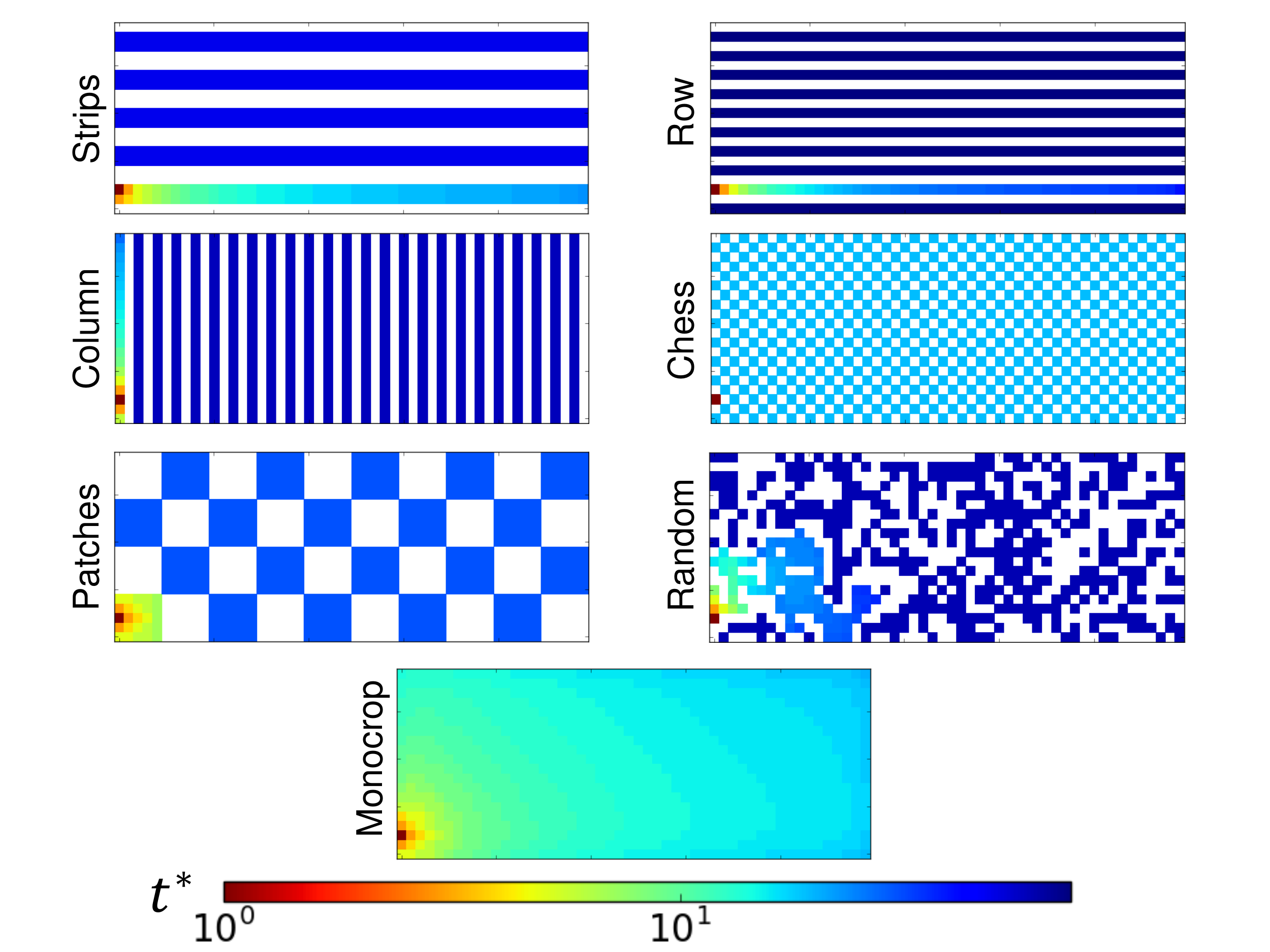}}
\par\end{centering}
\textcolor{black}{}}\textcolor{black}{}\subfloat[]{\begin{centering}
\textcolor{black}{\includegraphics[width=0.47\textwidth]{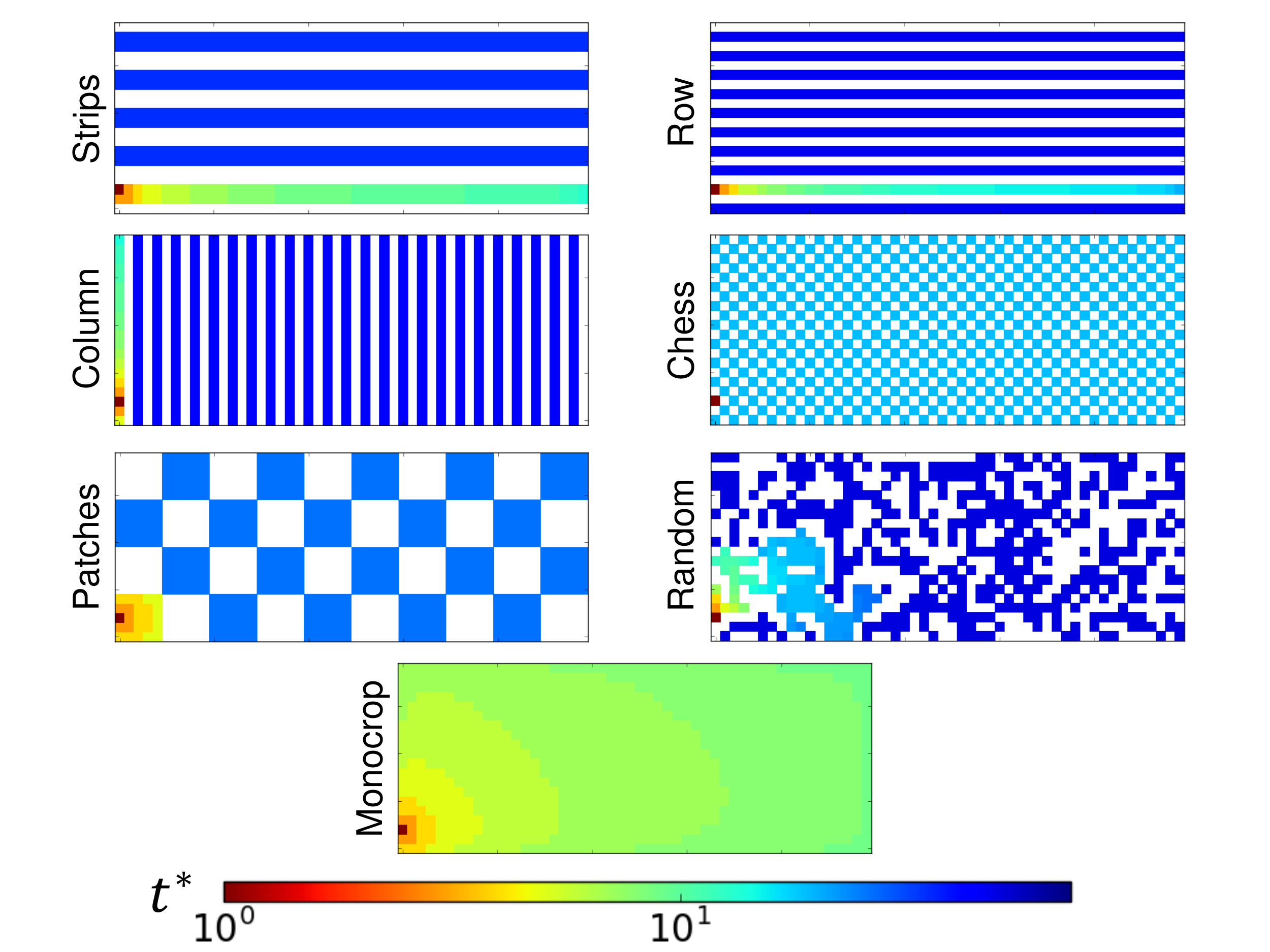}}
\par\end{centering}
\textcolor{black}{}}
\par\end{centering}
\caption{Raincloud plots of the proportion of dead plants under a SIR epidemics
at $t=10$ with $\beta=0.5,$ $\mu=0.5$ for different intercropping
strategies with perfect biological traps ($\gamma\rightarrow\infty$).
(a) Pest with reduced mobility ($s=4.0)$ and with larger mobility
($s=2.5$) The clouds show the kernel distribution of the proportion
of dead plants for different realizations of the epidemics. Below,
the raw data is plotted (the rain) together with their corresponding
box and whisker plots. Evolution of the propagation of a relatively
low mobility pest ($s=4.0)$ (c) and of a relatively high mobility
pest ($s=2.5)$ (d) across the fields. In both cases the propagation
is initialized by infecting the plant on the bottom-right corner of
the plot.}

\label{Traps-1}
\end{figure}

\section*{\textcolor{black}{Supplementary Note 5}}

\textcolor{black}{The value of the epidemic threshold $\tau$ can
be obtained as the reciprocal of largest eigenvalue $\lambda_{1}\left(G\right)$
of the adjacency matrix of the network representing the arrangement:}

\textcolor{black}{
\begin{equation}
\tau^{\,E}=\dfrac{1}{\lambda_{1}\left(G\right)},\label{eq:epidemic threshold}
\end{equation}
}

\textcolor{black}{Then, let $G$ be a field arrangement such that
in the shortest path connecting any two arbitrary pair of susceptible
plants $v_{i}$ and $v_{j}$ there is at least one trap crop. This
arrangement is a subdivision of certain network $G'$ in which a trap
crop is inserted between every pair of susceptible plants. Then, let
us consider the representation $G'$ in which only the susceptible
plants are represented. Every pair of susceptible plants in $G'$
are connected by a weighted edge with a weight $\left(2d\right)^{-\gamma s}$
where $d$ is the separation, in terms of number of edges, between
the two susceptible plants in the original arrangement $G$. The network
$G'$ is a weighted complete graph, in which every pair of nodes is
connected by a weighted edge. Thus, if $\left(\gamma s\right)\rightarrow0$,
which is realizable having $\gamma<\infty$ and an insect pest with
extremely large mobility $s\rightarrow0$, we have that $G'$ is a
graph formed by $n/2$ nodes in which every pair of nodes is connected
by an edge of weight equal to one. Consequently, the adjacency matrix
is $A\left(G'\right)=E-I$ where $E$ is the all-ones matrix and $I$
is the corresponding identity matrix. The largest eigenvalue of this
matrix is $\lambda_{1}\left(G'\right)=\dfrac{n}{2}-1.$Thus we have}

\textcolor{black}{
\begin{equation}
\tau^{\,E}=\underset{n\rightarrow\infty}{\lim}\underset{\left(\gamma s\right)\rightarrow0}{\lim}\dfrac{1}{\lambda_{1}\left(G'\right)}=\underset{n\rightarrow\infty}{\lim}\dfrac{1}{\dfrac{n}{2}-1}=0.
\end{equation}
}

\textcolor{black}{That is, in very large arrangements of this type
there are no epidemic threshold when the pest mobility is extremely
large and the trap crop has a fixed but not infinite trap strength.
This means that in this conditions just one single infestation of
a plant can trigger an epidemic in the crop field.}

\textcolor{black}{On the other hand, let us consider that $\left(\gamma s\right)\rightarrow\infty$,
which is realizable if $s>0$ and the trap crop has an extremely high
strength $\gamma\rightarrow\infty$. In this case every entry of the
adjacency matrix tends to zero $\left(2d\right)^{-\gamma s}\rightarrow0$
which means that the graph representing the system is formed by $n/2$
isolated nodes. In this case we have }

\textcolor{black}{
\begin{equation}
\tau^{\,E}=\underset{\left(\gamma s\right)\rightarrow\infty}{\lim}\dfrac{1}{\lambda_{1}\left(G'\right)}=\dfrac{1}{0^{+}}=\infty.
\end{equation}
}

\textcolor{black}{That is, in this ideal type of intercrop arrangement
in which there is one trap crop between every pair of susceptible
plants it is needed to infestate an extremely large number of plants
to trigger an epidemic if the strength of the trap is very high.}

\textcolor{black}{The previously studied 'ideal' intercropping system
is realizable by using the chessboard arrangement. That is, let us
build a chessboard arrangement in which the separation between every
pair of susceptible plants is larger than the radius that the insect
pest has for short range dispersal. Then, we will be in a situation
identical to that described by the ideal arrangement (see Supplementary
Figure \ref{SI_threshold}).}

\begin{figure}[H]
\begin{centering}
\textcolor{black}{\includegraphics[width=0.5\textwidth]{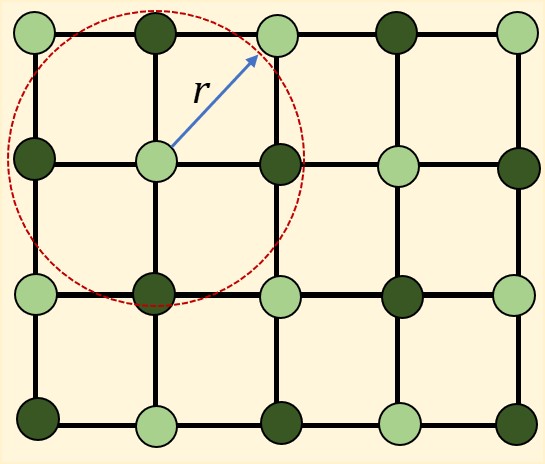}}
\par\end{centering}
\caption{Realization of an ideal intercrop system for a crop susceptible to
a pest (light green nodes) with trap crops (dark green nodes) for
having extremely large epidemic threshold when the strength of the
trap is very high.}

\label{SI_threshold}
\end{figure}

\end{document}